\documentclass[%
 preprint,
 amsmath,amssymb,
 aps,
]{revtex4-2}
\newcommand{\be}{\begin{equation}}
\newcommand{\ee}{  \end{equation}}
\newcommand{\ba}{\begin{eqnarray}}
\newcommand{\ea}{  \end{eqnarray}}

\usepackage{graphicx}% Include figure files
\usepackage{float}
\usepackage[dvipsnames]{xcolor}
\usepackage{bbm}
\usepackage{amsmath}
\usepackage{bbold}
\usepackage{mathtools}
\usepackage{dcolumn}% Align table columns on decimal point

\usepackage{subfig}

\usepackage{bm}% bold math
\usepackage{mathtools}
\usepackage{multirow}
\usepackage[colorlinks=true,linkcolor=blue,citecolor=blue,urlcolor=blue]{hyperref}% add hypertext capabilities

\usepackage{cleveref}

\RequirePackage[normalem]{ulem}
\usepackage{color}

\newcommand{\new}[1]{{\textcolor{blue}{#1}}}

\begin{document}

% \preprint{APS/123-QED}

\title{
{Eigenvalue Distribution of Empirical Correlation Matrices for Multiscale Complex Systems and Application to Financial Data}\\
}% Force line breaks with \\

\author{Luan M. T. de Moraes}
\affiliation{Laborat\'orio de F\'isica Te\'orica e Computacional, Departamento de F\'isica, Universidade Federal de Pernambuco, Recife, 50670-901, PE, Brazil}
\author{Ant\^onio M. S. Mac\^edo}
\affiliation{Laborat\'orio de F\'isica Te\'orica e Computacional, Departamento de F\'isica, Universidade Federal de Pernambuco, Recife, 50670-901, PE, Brazil}
\author{Giovani L. Vasconcelos}
\affiliation{Departamento de F\'isica, Universidade Federal do Paran\'a, Curitiba, 81531-980, PR, Brazil}
\author{Raydonal Ospina}
\affiliation{Departamento de Estat\'istica, Universidade Federal da Bahia, Salvador,  40170-110, BA, Brazil}
\affiliation{Departamento de Estat\'istica, Universidade Federal do Pernambuco,  Recife, 50670-901, PE, Brazil}

\date{\today}% It is always \today, today,

\begin{abstract}
We introduce a method for describing eigenvalue distributions of correlation matrices from multidimensional time series. Using our newly developed matrix H theory, we improve the description of eigenvalue spectra for empirical correlation matrices in multivariate financial data by considering an informational cascade modeled as a hierarchical structure akin to the Kolmogorov statistical theory of turbulence. Our approach extends the Marchenko-Pastur distribution to account for distinct characteristic scales, capturing a larger fraction of data variance, and challenging the traditional view of noise-dressed financial markets. We conjecture that the effectiveness of our method stems from the increased complexity in financial markets, reflected by new characteristic scales and the growth of computational trading. These findings not only support the turbulent market hypothesis as a source of noise but also provide a practical framework for noise reduction in empirical correlation matrices, enhancing the inference of true market correlations between assets.
% \begin{description}
% Secondary publications and information retrieval purposes.
% You may use the \texttt{description} environment to structure your abstract;
% use the optional argument of the \verb+\item+ command to give the category of each item. 
% \end{description}
\end{abstract}

\keywords{Random Matrix Theory, Econophysics, Eigenvalue Spectrum, Hierarchical Models, Financial Time Series, Stochastic Volatility }%Use showkeys class option if keyword
\maketitle

\section{Introduction}
\label{intro}

Physical systems can have multiple degrees of freedom, which can be recorded as multivariate time series. The environment coupled with the system works as a source of noise that manifests itself in the system's time series. \cite{strogatz2024nonlinear, van1992stochastic} Another source of noise comes with measurement, which may introduce some additional and uncontrollable disturbances, whether due to systematic measurement problems or the limited size of the available data. This additional noise  needs, therefore, to be separated from the true information we want to access. Researchers have sought to characterize noise and develop techniques to filter it from the relevant data. \cite{veraart2016denoising, veraart2016diffusion, vinayak2013emerging, tumminello2007shrinkage}

The empirical correlation matrix of a multivariate time series can be far different from its global true correlations due to noise or finite size effects. The data is then said to be noise dressed \cite{laloux1999noise}. Some of the eigenvalues of the correlation matrix might be associated with randomness, while  others are associated with true correlations in the data.
 In general, the eigenvalues associated with noise (called noisy eigenvalues here) are gathered inside a blob in the eigenvalues density, i.e., the bulk region of the eigenvalue spectrum. It is possible, in principle, to identify them by analyzing their distribution. \cite{plerou1999universal, plerou2002random, laloux1999noise}

Financial markets provide a large amount of data every year, and with advances in computation it has become easy to access and analyze them. Scientists have then tried to understand the basic mechanisms of the evolution of asset prices and their correlations \cite{vasconcelos2004guided, malkiel1999random, bouchaud2003theory}. Besides their potential uses for devising trading strategies and risk management tools, theoretical and empirical analyses of financial data have attracted physicists because financial markets are complex systems where concepts of statistical mechanics can be put to test \cite{voit2003statistical, mantegna1999introduction}. An old problem in statistical mechanics is the appropriate description of fully developed turbulence in fluids, which started with the studies of Kolmogorov and Obukhov in 1941 \cite{kolmogorov1941dissipation, obukhov1941distribution}. Later, Ghashghaie {\it et al.}~published a paper \cite{ghashghaie1996turbulent} that showed similar signatures of statistical turbulence in foreign exchange markets markets. Subsequent works further explored this connection \cite{arneodo1998direct, muzy2001multifractal}. In the early 2010's, Salazar and Vasconcelos \cite{salazar2010,salazar2012multicanonical} introduced a hierarchical stochastic model of intermittency containing key ingredients  of statistical turbulence to describe multiscale complex systems, such as fluid turbulence and financial markets.  Macêdo {\it et al.}~\cite{macedo2017universality} subsequently extended the Salazar-Vasconcelos model  into a more general framework  that  became known as H theory. The H-theory formalism aims to explain the intermittent behavior of certain physical signals (such as velocity increments in fluid turbulence or price fluctuations in financial markets) by assuming that the system evolves under a multiscale fluctuating background (for example,  the energy transfer rate between turbulent eddies or  price volatility at different time scales). These authors showed that the background hierarchical   dynamics generally falls into two universality classes  \cite{macedo2017universality}. 
%In the same paper, the authors managed to use the same model to explain the probability distributions of the increments of velocities in a turbulent fluid and the financial returns of the Ibovespa Index. Later, 
Recently, Vasconcelos {\it et al.}~\cite{vasconcelos2024turbulence} used the H-theory approach to give more evidence that foreign exchange markets indeed have strong signatures of turbulence, thus corroborating the results by Ghashghaie {\it et al.~}\cite{ghashghaie1996turbulent}. And more recently,  the present authors published a study \cite{de2025matrix} 
where a multivariate form of H theory, dubbed matrix H theory,  was developed with the aid of random matrix theory (RMT), with applications 
 to  financial time series from the S\&P 500 stock index. 

The extension of H theory to multivariate systems leads to new questions, since multivariate time series displays correlations among components that can be reflected in the statistics of the empirical correlation matrix and in its eigenvalue spectrum. Plerou {\it et al.} \cite{plerou1999universal, plerou2002random} reported several universal and non-universal properties of correlation matrices in financial time series as well as their eigenvectors and eigenvalues statistics. Analyzing data from the 1990's, these studies found that the Marchenko-Pastur (MP) distribution fitted reasonably well the eigenvalue spectrum of the correlation matrix of financial returns of S\&P 500 index.  Their findings, however, 
do not appear to hold for more recent  financial data. As we report later in this paper, the  MP distribution does not fit the eigenvalue spectrum of the correlation matrix  of the S\&P 500 returns after 2020. The hierarchical  behavior in financial markets predicted by the turbulence analogy considers an asymmetry in the information flow between long- and short-term investors \cite{vasconcelos2024turbulence, zumbach2001heterogeneous, muller1997volatilities, chakrabarty2015investment, genccay2010asymmetry}. Therefore, it is reasonable to expect that this
 cascade-like dynamics may account for the aforementioned discrepancies in the eigenvalue spectra of correlation matrices of financial time series,  where the eigenvalue distribution for more recent data---and presumably more turbulent-like---cannot be described by the standard MP distribution. 

In this article, we introduce a generalized version of the MP distribution for a hierarchical multivariate process considering the two universality classes predicted by matrix H theory. We demonstrate that these new distributions provide significantly better fits when accounting for the number of relevant time scales, consistent with our previous work \cite{de2025matrix}. The results presented here show that noise-dressed data is better described by considering its hierarchical behavior, corroborating the turbulent market hypothesis. Furthermore, this approach can potentially separate eigenvalues related to noise from those corresponding to actual correlations in market data, making it valuable for portfolio management applications.

This paper is organized as follows. Section \ref{sec:review} reviews the basic concepts of matrix H theory. Sec.~\ref{sec:eigen} introduces the MP distribution and generalizes it considering the hierarchical structure of   H theory, while Sec.~\ref{applications} applies the theory developed here to  financial returns of the S\&P 500 stock index. 
%In Section \ref{HMT} is made an analysis of the same dataset but in terms of the analysis perform in \cite{de2025matrix} to choose the most appropriate model in Section \ref{applications}
In Sec.~\ref{discussions} we discuss our main results, and  conclusions are presented in Sec.~\ref{conclusions}.

\section{Matrix H Theory: Brief review} 
\label{sec:review}

In this section, we briefly review the multivariate version of the H-theory formalism,  to render the paper as self-contained as possible. For more details (on both the univariate and multivariate  H theory) the reader should refer to Refs.~\cite{salazar2012multicanonical, macedo2017universality, de2025matrix}.

Let $\boldsymbol{r}^\top=(r_1,r_2,...,r_p)$ be a   random vector in $\mathbbm{R}^p,$ where the superscript $\top$ stands for transpose. For example, each random variable $r_i(t)$ may represent the returns of a given stock computed at some short time scale $\tau_N$ (say, daily or intraday returns). The set of $p$ companies considered may correspond, for instance, to  companies from a given sector of the economy or companies that enter a given stock exchange index. 
To be specific, we assume that there is a number,  $N$, of well-separated times scale, $\tau_i$,  between the shortest time scale, $\tau_N$, 
%where the measurements  are made, 
and the largest time scale, $\tau_0$, above which  no correlations in the series would be present, so that  $\tau_{N} 
%\ll \tau_{N-1}
\ll \cdots \ll \tau_{1}\ll \tau_{0}$. 
The large time-scale multivariate distribution of returns,   $P_0(\boldsymbol{r}|\Sigma_0)$,  is  assumed to be known, where the  parameter matrix $\Sigma_0 $ characterizes the global (large-scale) `equilibrium' of the system. 
More precisely, on account of the central limit theorem we  assume  that $P_0(\boldsymbol{r}|\Sigma_0)$ is  a multivariate Gaussian distribution:
\begin{equation}
\label{Gauss2}
P_0(\boldsymbol{r}|\Sigma_0)=\frac{1}{|2\pi\Sigma_0|^{1/2}}\exp\left( -\frac{1}{2}\boldsymbol{r}^\top\Sigma_0^{-1}\boldsymbol{r}\right),
\end{equation}
where $\Sigma_0$ is the large-scale correlation matrix and $|\Sigma_0|\equiv\det(\Sigma_0)$. 
%and all variances can be set to be the same.

The first  main principle of matrix H theory \cite{macedo2017universality} is  that the short scale distribution $P_N(\boldsymbol{r})$  is obtained from a compound of the large-scale distribution with a weighing distribution that describes the fluctuating local (short-scale) `background':
\begin{equation}\label{signal2}
P_N(\boldsymbol{r})=\int P_0(\boldsymbol{r}|\Sigma_N)f_N(\Sigma_N)d\Sigma_N,
\end{equation}
where $\Sigma_N$ is the short-scale correlation matrix
and $f_N(\Sigma_N)$, its probability density.  Notice that in Eq.~(\ref{signal2}) the integral is over the space of $p \times p$ real symmetric semi-positive definite matrices $\Sigma_N$ (i.e., an integral over all of its entries).  The physical idea captured in (\ref{signal2}) is that over short periods of times the system tends to relax toward a quasi-equilibrium described by the same functional-form distribution as in the large-scale equilibrium. But, now, the local equilibrium parameters, characterizing the fluctuating background, are  given by a random matrix $\Sigma_N$, indicating that the variances and correlations of the signals $r_i(t)$ slowly fluctuate in time \cite{forrester2010log}. 

The second main principle of the H-theory approach is that the multivariate background probability density $f_N(\Sigma_N)$ is obtained from a hierarchical series of convolutions:
\begin{align}
\label{background2}
f_N(\Sigma_N)=\int  d\Sigma_1 ...d\Sigma_{N-1} \prod_{i=1}^{N}f(\Sigma_i|\Sigma_{i-1}),  
\end{align}
where $f(\Sigma_i|\Sigma_{i-1})$ is the background density at scale $\tau_i$ for fixed $\Sigma_{i-1}$, with the property that  $\langle \Sigma_i|\Sigma_{i-1}\rangle=\Sigma_{i-1}$.  This choice is motivated by the fact that information propagates from large  to small  time scales \cite{vasconcelos2024turbulence}. Furthermore, we require that $\langle \Sigma_i\rangle=\Sigma_{0}$, $i=1,...,N$, to reflect the average global equilibrium of the system.
Under rather general arguments, one can show \cite{de2025matrix} that there are two physically relevant choices for the background density $f(\Sigma_i|\Sigma_{i-1})$, namely: (i) Wishart and (ii) inverse-Wishart distributions, which are   the multivariate extensions of the gamma and inverse-gamma distributions, respectively. 
Let us  consider these two classes  separately.

\subsection{Wishart class}

For the Wishart class, one takes  $f(\Sigma_i|\Sigma_{i-1})$ to be a Wishart distribution of the form
\begin{align}
\label{wishartdistribution}
f(\Sigma_i|\Sigma_{i-1})=&\frac{|\beta_i\Sigma_{i-1}^{-1}|^{\beta_i}}{\Gamma_p(\beta_i)}|\Sigma_i|^{\beta_i-(p+1)/2} \nonumber \\ & \times  \; \exp\left(-\beta_i{\rm Tr}\left( \Sigma_{i-1}^{-1}\Sigma_i\right)\right),
\end{align} 
where the $\beta_i$'s are positive parameters and 
\begin{align}
\label{multivariategamma}
\Gamma_p(\beta)\equiv\pi^{p(p-1)/4}\Gamma(\beta)\Gamma(\beta-\tfrac{1}{2})\Gamma(\beta-1)\cdots\Gamma(\beta-\tfrac{(p-1)}{2}),  
\end{align} 
is the matrix-variate gamma function \cite{mathai1991multivariate}. 
Inserting (\ref{wishartdistribution}) into (\ref{background2}), and using  properties of the Mellin transform, one finds  
\begin{equation}
\label{gamma2}
    f_N(\Sigma_N )=
\frac{|\omega \Sigma_0^{-1}|^{(p+1)/2}}{\Gamma_p(\boldsymbol \beta)}   \bar{G}_{ 0,N } ^{ N,0 }  \left( 
\begin{array}{c}
{-} \\ 
{ \boldsymbol\beta-\frac{p+1}{2}{\bf 1}}
\end{array}
\bigg |\frac{\omega \Sigma_N}{\Sigma_0 }  \right),
\end{equation}
where  $\omega  =\prod_{j=1}^{N}\beta _j$ and we have introduced the vector notations
${\boldsymbol\beta}\equiv (\beta_1,\dots,\beta_N)$ and 
$
\Gamma({\bf a})  \equiv\prod_{j=1}^{N}\Gamma (a _j)$. Here $\bar G$  denotes the matrix-argument Meijer $G$ function.
%, so as to avoid confusion with the standard one-variable $G$ function.
%
%

Now, inserting (\ref{gamma2}) into (\ref{signal2}) 
%and performing a change of variables, $X=(\omega \Sigma_0^{-1})^{1/2}\Sigma_N(\omega \Sigma_0^{-1})^{1/2}$,
one obtains
\begin{align}
\label{CFT1}
P_N(\boldsymbol{r})=&\frac{|\omega \Sigma^{-1}_0|^{(p+1)/2}}{\Gamma_p(\boldsymbol \beta)}\int d\Sigma_N P_0\left(\boldsymbol{r}|\Sigma_N\right) \nonumber\\ &\times\;  \bar{G}_{ 0,N } ^{ N,0 }  \left( 
\begin{array}{c}
{-} \\ 
{ \boldsymbol\beta-\frac{p+1}{2}{\bf 1}}
\end{array}
\bigg |\frac{\Sigma_N \omega}{\Sigma_0} \right).
\end{align}
The difficulty with this expression is that it involves a matrix integration.  To circumvent this, one can use a color-flavor-type transformation (CFT), which maps the multivariate integral above to its univariate version \cite{zirnbauer2021color, wei2009n, forrester2010log}. After such a procedure (see \cite{de2025matrix} for details), Eq.~(\ref{CFT1}) becomes 
\begin{align}
\label{CFT1b}   
    P_N({\bf r}) 
    =& \frac{1}{\Gamma({\bf \boldsymbol{\beta}})} \displaystyle \int dx P_0\left({\bf r}\middle| \frac{x}{\omega} \Sigma_0 \right) G^{N, 0}_{0, N} \left( 
    \begin{array}{c}
   -\\
    \boldsymbol{\beta} - {\bf 1}
    \end{array} \middle \vert x
    \right) ,
\end{align}
where the integral is now one-dimensional and  $G$  denotes the standard one-variable Meijer $G$ function. 
Using  properties of the $G$ function, the integral can be readily computed to yield:
\begin{equation}
P_N(\boldsymbol{r})=
\frac{\omega^{p/2}}{{|2\pi\Sigma_0|^{1/2}}\Gamma(\boldsymbol\beta)}G_{0,N+1}^{N+1,0}\left( 
\begin{array}{c}
- \\ 
{ \boldsymbol\beta-\frac{p}{2}{\bf 1}},0
\end{array}
\bigg |\frac{\omega}{2}\boldsymbol{r}^\top\Sigma_0^{-1}\boldsymbol{r}\right).
\label{eq:PN3}
\end{equation}

\subsection{Inverse Wishart class}

For the inverse-Wishart class of hierarchical matrix distributions one chooses for $f(\Sigma_i | \Sigma_{i-1})$  an inverse-Wishart distribution, 
\begin{align}
 \label{inverseWishart}
    f(\Sigma_i | \Sigma_{i-1}) =~& \frac{|\beta_i \Sigma_{i-1}|^{\beta_i + (p+1)/2}}{\Gamma_p (\beta_i + \frac{p+1}{2})} |\Sigma_i|^{-\beta_i -p-1} \nonumber \\ &\times \exp{(-\beta_i \text{Tr}(\Sigma_i^{-1} \Sigma_{i-1}))},
\end{align}
which upon insertion into  (\ref{background2}) yields
\begin{equation}
    \label{inverse}
    f_N(\Sigma_N) = \frac{|\omega \Sigma_0|^{-(p+1)/2}}{\Gamma_p( \boldsymbol{\beta} + \frac{p+1}{2}{\bf 1})} \overline{G}^{0, N}_{N, 0} \left( 
    \begin{array}{c}
   -\boldsymbol{\beta} - \frac{p+1}{2}{\bf 1}\\
    -
    \end{array} \middle \vert \frac{\Sigma_N}{\omega \Sigma_0}
    \right).
\end{equation}
The compound integral (\ref{signal2}) in this case gives
\begin{align}
\label{CFT2}
    P_N({\bf r}) =& \frac{|\omega \Sigma_0|^{-(p+1)/2}}{\Gamma_p( \boldsymbol{\beta} + \frac{p+1}{2}{\bf 1})} \int d\Sigma_N P_0({\bf r} |\Sigma_N)  \nonumber  \\ &\times\; \overline{G}^{0, N}_{N, 0} \left( 
    \begin{array}{c}
   -\boldsymbol{\beta} - \frac{p+1}{2}{\bf 1}\\
    -
    \end{array} \middle \vert \frac{\Sigma_N}{\omega\Sigma_0}
    \right) .
\end{align}
Performing again a CFT \cite{de2025matrix}, one then finds a  one-dimensional integral
\begin{align}
\label{CFT2b}
    P_N({\bf r}) = \frac{1}{\Gamma({\bf \boldsymbol{\beta}} +{\bf \boldsymbol{1}})} \displaystyle \int dx P_0\left({\bf r}| x \omega \Sigma_0 \right) G^{0, N}_{N, 0} \left( 
    \begin{array}{c}
   -\boldsymbol{\beta} - {\bf 1}\\
    -
    \end{array} \middle \vert x
    \right) ,
\end{align}
which can be computed exactly to give \cite{de2025matrix} 
%(on using  properties of the $G$ function) and performing the integral one obtains 

%
\begin{equation}
\label{inverse_wishart}
    P_N({\bf r}) = \frac{| 2 \pi \Sigma_0|^{-1/2}}{\omega^{p/2}\Gamma({\bf \boldsymbol{\beta}} + {\bf \boldsymbol{1}})} {G}^{1, N}_{N, 1} \left( 
    \begin{array}{c}
   -\boldsymbol{\beta} - \frac{p}{2}{\bf 1}\\
    0
    \end{array} \middle \vert \frac{{\bf r}^\top \Sigma_0^{-1} {\bf r}}{2\omega}
    \right).
\end{equation}
\subsection{Univariate distribution of returns}

Let us assume that, over long time separations and upon proper normalization, the components of the multivariate process ${\bf r}(t)$ are all comparable and statistically indistinguishable from each other. This implies that the large scale correlation matrix $\Sigma_0$ matrix can be taken as a multiple of the identity matrix: $\Sigma_0 = \varepsilon_0 \mathbb{1} $, where $\varepsilon_0$ is the common variance of the series.
Performing a change of variables,  we can rewrite the scalar integrals  (\ref{CFT1b}) and (\ref{CFT2b})  as a one-dimensional compound: 
\begin{equation}
    P_N({\bf r}) = \int_0^{\infty} d\varepsilon_N  \frac{\exp (-{\bf r}^\top {\bf r}/2\varepsilon_N)}{(2 \pi \varepsilon_N )^{1/2}}f_N(\varepsilon_N),
    \label{eq:compound}
\end{equation}
where  $f_N(\varepsilon_N)$ is the equivalent univariate background distribution given by
\begin{equation}
    f_N(\varepsilon_N) = \frac{\omega}{ \varepsilon_0\Gamma(\boldsymbol \beta)}\;G_{ 0,N } ^{ N,0 }  \left( 
\begin{array}{c}
{-} \\ 
{ \boldsymbol\beta-{\bf 1}}
\end{array}
\bigg |\frac{\omega\varepsilon_N}{\varepsilon_0}   \right ).
\label{eq:Wback}
\end{equation}
for the Wishart class, and 
\begin{equation}
    f_N(\varepsilon_N) = \frac{1}{\omega \varepsilon_0\Gamma({\bf \boldsymbol{\beta}} + {\bf \boldsymbol{1}})}\; G^{0, N}_{N, 0} \left( 
    \begin{array}{c}
   -\boldsymbol{\beta} - {\bf 1}\\
    -
    \end{array} \middle \vert \frac{\varepsilon_N}{ \omega\varepsilon_0} 
    \right),
    \label{eq:invWback}
\end{equation}
for the Inverse-Wishart class. 
We shall thus refer to the univariate distributions $f_N(\varepsilon_N)$ in (\ref{eq:Wback}) and (\ref{eq:invWback}) as the one-dimensional projection of the multivariate distributions $f_N(\Sigma_N)$ given in (\ref{gamma2}) and (\ref{inverse}), respectively. We note that, by construction, $\langle \varepsilon_N\rangle = \varepsilon_0$ for both classes of background distributions \cite{de2025matrix}.

For later use, we note that  upon using the asymptotic expansion for the $G$ functions one finds \cite{sosa2018hierarchical} that  $f_N(\varepsilon)$ in (\ref{eq:Wback}) has a modified stretched-exponential tail (for $\varepsilon_N\to \infty$), 
\begin{equation}
    f_N(\varepsilon_N) \propto 
\left(\frac{\omega \varepsilon}{\varepsilon_0}\right)^{\beta - \frac{3}{2} + \frac{1}{2N}} \exp \left[ -\beta N\left(\frac{\varepsilon}{\varepsilon_0} \right)^{1/N}\right];
\label{eq:tailFW}
\end{equation}
whereas the distribution in  (\ref{eq:invWback}) has a power-law tail for ($\varepsilon_N \rightarrow \infty$),

\begin{equation}
    f_N(\varepsilon_N) \propto \left( \frac{\varepsilon_N}{\varepsilon_0 \omega}\right)^{-\beta - 2}
    \label{eq:tailfIW}
\end{equation}
where in both cases we have set $\beta_i = \beta$.

The distribution $P_N({\bf r})$  is now symmetric under permutation of components of ${\bf r}$. Let $\tilde{r}$ be one given component of ${\bf r}$. If we integrate $P_N({\bf r})$ over all other components (i.e., ${\bf r}$ except component $\tilde{r}$), we obtain a univariate projection distribution of ${\bf r}$:
\begin{equation}
    P_N(\tilde{r}) = \int_0^{\infty} d\varepsilon_N  \frac{\exp (-\tilde{r}^2/2\varepsilon_N)}{\sqrt{2 \pi \varepsilon_N}}f_N(\varepsilon_N).
    \label{eq:univariatecompound}
\end{equation}
For $f_N(\varepsilon_N)$  given by Eqs.~(\ref{eq:Wback}) and (\ref{eq:invWback}), the integral in Eq.~(\ref{eq:univariatecompound}) equals respectively to Eqs.~(\ref{eq:PN3}) and (\ref{inverse_wishart}) with $p = 1$. More specifically, for the Wishart class we have
\begin{align}
\label{wishartaggregated}
P_N(\tilde{r})=
\frac{\omega^{1/2}}{\sqrt{2\pi \varepsilon_0}\Gamma(\boldsymbol\beta)}G_{0,N+1}^{N+1,0}\left( 
\begin{array}{c}
- \\ 
{ \boldsymbol\beta-{\bf 1/2}},0
\end{array}
\bigg |\frac{\omega \tilde{r}^2}{2\varepsilon_0}\right),
\end{align}
whilst for the inverse-Wishart class one obtains 
\begin{align}
\label{inversewishartaggregated}
    P_N(\tilde{r})=
\frac{(2\pi \omega 
\varepsilon_0)^{-1/2}}{\Gamma({\bf \boldsymbol{\beta}} + {\bf \boldsymbol{1}})}G_{N,1}^{1,N}\left( 
\begin{array}{c}
{ -\boldsymbol{\beta}- \frac{1}{2}{\bf 1}} \\ 
0
\end{array}
\bigg |\frac{\tilde{r}^2}{2\omega \varepsilon_0}\right). 
\end{align}

Summarizing this section, we have indicated above how to obtain, within the matrix H-theory formalism,  two classes of hierarchical multivariate distributions. One important step in the derivations  was the use of a CFT that allowed us to recast certain matrix integrals, namely Eqs.~(\ref{CFT1}) and (\ref{CFT2}), in terms of their one-dimensional counterparts, Eqs.~(\ref{CFT1b}) and (\ref{CFT2b}).
%which then allowed us to obtain analytical expressions for both the Wishart and inverse Wishart classes of distributions, as shown in Eqs.~(\ref{eq:PN3}) and (\ref{inverse_wishart}), respectively. The CFT property will also be important in deriving the distribution of eigenvalues of the correlation matrix of multivariate returns, as discussed next.
%
More importantly for what follows, the CFT property allowed us to  establish a connection between the multivariate background distributions  $f(\Sigma_N)$ and their  univariate counterparts $f(\varepsilon_N)$. 
%We thus  refer  to the latter as a one-dimensional projection  of the former. 
For convenience, we show  in table \ref{tab:comparison} the relations between the multivariate background distributions and their univariate projections for both the Wishart and inverse Wishart classes.
This correspondence shows, in turn, that the distribution $P_N({\bf r})$ can be obtained from two different procedures, namely: (i) the {\it matrix method}, where one performs a compounding  of the conditional distribution $P_0({\bf r}|\Sigma_N)$ with a matrix weighing  distribution  $f_N(\Sigma_N)$; and (ii)  the {\it scalar  method}, where one compounds the conditional   distribution $P_0({\bf r}|\mathbb{1}\varepsilon_N)$ with a   univariate weighing distribution, $f_N(\varepsilon_N)$. In other words, we can write 
\begin{equation}
    \int P({\bf r}|\Sigma_N) f_N(\Sigma_N) d\Sigma_N = \int P({\bf r}|\mathbb{1}\varepsilon_N)f_N(\varepsilon_N)d\varepsilon_N.
    \label{eq:CFT}
\end{equation}
Thus, the CFT claims that random vectors ${\bf r}$ can be sampled from $P_N({\bf r})$ in two different ways. This property will be used below to derive the distribution of eigenvalues of empirical correlation matrices for  random vectors described by either (\ref{eq:PN3}) or (\ref{inverse_wishart}).

\begin{table*}
    \centering
    \renewcommand{\arraystretch}{1.5}
    \begin{tabular}{c|c|c}
       & Matrix Distribution  & (Projected) Univariate Distribution \\ \hline
       $s = 1$ & \parbox{7cm}{\centering 
       $\frac{\omega^{p(p+1)/2}}{\varepsilon_0\Gamma_p(\boldsymbol \beta)}
       \bar{G}_{ 0,N } ^{ N,0 }  \left( 
       \begin{array}{c}
       {-} \\ 
       { \boldsymbol\beta-\frac{p+1}{2}{\bf 1}}
       \end{array}
       \bigg |\frac{\omega \Sigma_N}{\varepsilon_0} \right)$} 
       & 
       \parbox{7cm}{\centering 
       $\frac{\omega}{\varepsilon_0 \Gamma(\boldsymbol \beta)} 
       G_{ 0,N } ^{ N,0 }  \left( 
       \begin{array}{c}
       {-} \\ 
       { \boldsymbol\beta-{\bf 1}}
       \end{array}
       \bigg |\frac{\omega \varepsilon_N}{\varepsilon_0} \right)$} \\ \hline
       $s = \frac{1}{2}$ & \parbox{7cm}{\centering 
       $\frac{\omega^{-p(p+1)/2}}{\varepsilon_0\Gamma_p( \boldsymbol{\beta} + \frac{p+1}{2}{\bf 1})} \overline{G}^{0, N}_{N, 0} \left( 
    \begin{array}{c}
   -\boldsymbol{\beta} - \frac{p+1}{2}{\bf 1}\\
    -
    \end{array} \middle \vert \frac{\Sigma_N}{\omega \varepsilon_0}
    \right)$} 
       & 
       \parbox{7cm}{\centering 
       $\frac{1}{\varepsilon_0\omega\Gamma({\bf \boldsymbol{\beta}} + {\bf \boldsymbol{1}})} 
       G_{ N,0 } ^{ 0,N }  \left( 
       \begin{array}{c}
       {-\boldsymbol{\beta} - {\bf 1}} \\ 
       -
       \end{array}
       \bigg |\frac{ \varepsilon_N}{\omega \varepsilon_0} \right)$} \\ 
       \hline
    \end{tabular}
    \caption{The multivariate distributions for the matrix background of the Matrix H theory and the respective univariate projections which coincide with the background distributions for the univariate H theory.}
    \label{tab:comparison}
\end{table*}

\section{Eigenvalue distribution of empirical correlation matrices}\label{sec:eigen}

Let ${\boldsymbol \eta}_t$ be a Gaussian random vector with correlation matrix equal to $\mathbb{1}$ and let $\Sigma_t$ and $\varepsilon_t$ be respectively the random matrix and random variable sampled from their respective matrix and projected univariate background distributions, as indicated in table \ref{tab:comparison}. 
Then, in view of (\ref{eq:CFT}), the random vector ${\boldsymbol r}_t$ can be sampled in terms of both scalar and matrix backgrounds:
\begin{equation}
    {\boldsymbol r}_t \overset{d}{=} \Sigma_t^{1/2} {\boldsymbol \eta}_t \overset{d}{=} \varepsilon_t^{1/2} {\boldsymbol \eta}_t,
    \label{eq:CFT2_sec3}
\end{equation}
where $\overset{d}{=}$ represents {\it equality in distribution} and the subindex $t$  represents different sampling. In this case we say that $t$ plays the role of sampling time. 
Now, let us build a random time series of size $T$ where each point is sampled from distribution $P_N({\bf r})$. Thus we need to sample $T$ random vectors ${\boldsymbol r}_t$. Using both the scalar and matrix sampling methods will produce the same stationary distribution $P_N$, but it is clear that they have totally different correlation structures since $\varepsilon_t$ is the projection of the random matrix variable $\Sigma_t$. 

The empirical correlation matrix, $C$, of the random time series sampled from the random vectors ${\boldsymbol r}_t = \Sigma_t^{1/2} {\boldsymbol \eta}_t$
is
\begin{align}
C = \sum_t \frac{{\boldsymbol r}_t {\boldsymbol r}_t^\top}{T} = \sum_t \frac{\Sigma_t^{1/2} {\boldsymbol \eta}_t {\boldsymbol \eta}_t^\top\Sigma_t^{1/2}}{T} \equiv \sum_t P_t,
\label{eq:C_sec3}
\end{align}
where $P_t$ is a $p\times p$ matrix  defined by
\begin{equation}
        P_t = \frac{1}{T} \Sigma_t^{1/2} {\boldsymbol \eta}_t {\boldsymbol \eta}_t^\top \Sigma^{1/2}_t ,
\end{equation}        
which  is a projector-like operator, as shown next.
Let us calculate $P^2_{t}$:
\begin{align}
        P^2_t =& \frac{1}{T^2} \Sigma_t^{1/2} {\boldsymbol \eta}_t {\boldsymbol \eta}_t^\top \Sigma^{1/2}_t \Sigma^{1/2}_t {\boldsymbol \eta}_t {\boldsymbol \eta}_t^\top \Sigma_t^{1/2} \cr
        =&\frac{1}{T^2} \Sigma_t^{1/2} {\boldsymbol \eta}_t {\boldsymbol \eta}_t^\top\Sigma_t {\boldsymbol \eta}_t {\boldsymbol \eta}_t^\top \Sigma_t^{1/2} \cr
        \equiv& \frac{x_t}{T^2} \Sigma^{1/2}_t {\boldsymbol \eta}_t {\boldsymbol \eta}_t^\top \Sigma_t^{1/2} = \frac{x_t}{T}P_t,
    \label{eq:Pt2}
\end{align}
where 
\begin{align}
    x_t = {\boldsymbol \eta}_t^\top \Sigma_t {\boldsymbol \eta}_t.
    \label{eq:x1}
\end{align}
Now, note that
\begin{equation}
    x_t = {\boldsymbol \eta}_t^\top \Sigma_t {\boldsymbol \eta}_t \overset{d}{=} \varepsilon_t {\boldsymbol \eta}_t^\top{\boldsymbol \eta}_t \sim \varepsilon_t  p ,
    \label{eq:x2}
\end{equation}
where $\sim$ means {\it asymptotically equal} when $p$ is large \new{and} ${\boldsymbol \eta}_t^\top {\boldsymbol \eta_t}$ is the sum of $p$ independent identically distributed squared Gaussian random variables (law of large numbers). Defining 
\begin{align}
    q_t = \frac{x_t}{T}= \frac{p}{T}  \varepsilon_t = q \varepsilon_t ,
    \label{eq:qt}
\end{align} 
where $q=p/T$,
we obtain from (\ref{eq:Pt2}) that 
\begin{equation}\label{remarkbiroli}
    P_t^2 =  q_t P_t,
\end{equation}
thus showing that $P_t$ is indeed a projector-like operator.

Here we are interested in computing the 
eigenvalue spectral density (ESD), $\rho(\lambda)$, of the empirical correlation matrix $C$ for the cases where ${\boldsymbol r}_t$ are sampled according to the hierarchical distributions derived  in Sec.~II. To do that, we first recall that the ESD of $C$ is related to its resolvent, 
\begin{align}
g_C(z)= \frac{1}{p} \operatorname{Tr} \left( \frac{1}{zI - C} \right),
 \label{eq:gC}
\end{align}
via  the following formula (Stieltjes transform) \cite{potters2020first}:
\begin{align}
g_C(z) 
 = \int_{-\infty}^{\infty} \frac{\rho_C(\lambda)}{z - \lambda} \, d\lambda,
 \label{eq:glambda}
\end{align}
which is valid  in the limit of large $p$ and $T$. From (\ref{eq:glambda}), it then follows that
\begin{align}
\rho_C(\lambda) = \lim_{\varepsilon \to 0^+} \frac{1}{\pi} \operatorname{Im} \left[ g_C(\lambda + i \varepsilon)\right].
\label{eq:rho} 
\end{align}
So our task  is to compute the resolvent $g_C(z)$. But since $C$ is a sum of projectors, see (\ref{eq:C_sec3}),  we first need to compute the resolvent of $P_t$, as indicated next.

It is a standard calculation  to show that  for a projector as in (\ref{eq:C_sec3}) one has
\begin{equation} \label{resolvent}
   \frac{1}{z - P_t} = \frac{1}{z}  + \frac{P_t}{z(z-q_t)}.
\end{equation}
In view of definition (\ref{eq:gC}), it then follows that
\begin{equation}
\begin{split}
    g_P(z) 
    &= \frac{1}{z} + \frac{q_t}{pz(z-q_t)} ,
\end{split}
\label{eq:gP}
\end{equation}
where we have used that 
\begin{align}
    \text{tr} P_t &= \sum_i \frac{( \Sigma^{1/2}_t {\boldsymbol \eta}_t {\boldsymbol \eta}_t^\top \Sigma^{1/2}_t)_{ii}}{T} \\
    &= \frac{{\boldsymbol \eta}_t^\top \Sigma_t {\boldsymbol \eta}_t}{T} = \frac{x_t}{T} \sim q_t,
\end{align}
with the last identity following  from the definition of $q_t$ in  (\ref{eq:qt}).

In order to proceed we must calculate the inverse of the resolvent $z_P(g)$.
Setting $g_P(z)=g$ and solving (\ref{eq:gP}) for $z$ we obtain
\begin{equation}
    z_\pm = \frac{1 + \frac{1}{g} \pm \sqrt{\left(q_t + \frac{1}{g}\right)^2 - 4\frac{q_t}{g} \left( 1 - \frac{1}{p} \right)}}{2}
\end{equation}
Expanding the square root up to first order in $1/p$ we find
\begin{equation}
    z_\pm = \frac{q_t}{2} + \frac{1}{2g} \pm \left[ \frac{q_t}{2} - \frac{1}{2g} + \frac{q_t}{gp} \left( q_t - \frac{1}{g}\right)\right]
\end{equation}
When $z$ is large, we expect $g$ to be asymptotically equal to $1/z$ \cite{potters2020first}. So we take the negative root as the actual solution:
\begin{equation}
    z_P(g) 
    = \frac{1}{g} + \frac{1}{p}\left( \frac{q_t}{1 - g q_t} \right).
    \label{eq:zt}
\end{equation}

Next, we recall that the $R$-transform of an arbitrary random matrix ${\bf A}$ is written in terms of its inverse resolvent  $z_A(g)$ as
\begin{equation}
\label{Rtransform}
    R_{A} (g)  =  z_A(g)-\frac{1}{g}
\end{equation}
Denoting by $R_t(g)$ the $R$-transform of $P_t$ and using (\ref{eq:zt}), we  obtain
\begin{equation}
    R_t(g) = \frac{1}{p}\left( \frac{q_t}{1 - g q_t} \right).
\end{equation}
In the limit of large random matrices, the $R$ transform  is additive \cite{tulino2004random}. Then, the $R$-transform of $C = \sum_t P_t$ is simply the sum of the $R$-transforms of $P_t$: 
\begin{equation}
    \begin{split}
        R_C(g) &= \sum_t R_t(g) \\
        &= \sum_t \frac{1}{p}\left( \frac{q_t}{1 - g q_t} \right) \\
        &= T \sum_t \frac{\varepsilon_t }{1 - g q \varepsilon_t },
    \end{split}
\end{equation}
where in the last passage we used (\ref{eq:qt}).
In the limit of large $p$ and $T$, the sum in the last equality converges to an integral:
\begin{equation}
    R_C(g) = \int d\varepsilon f(\varepsilon) \frac{\varepsilon }{1 - qg \varepsilon },
    \label{eq:RC}
\end{equation}
where $f(\varepsilon)$ is the probability density of the background variable $\varepsilon$. 
From (\ref{eq:RC}) and  (\ref{Rtransform}), we thus obtain the inverse resolvent for the correlation matrix $C$:
\begin{equation} \label{blue}
    z(g) = \frac{1}{g} + \int d\varepsilon f(\varepsilon) \frac{\varepsilon }{1 - qg \varepsilon },
\end{equation}
where we have omitted the subscript $C$.
Notice that if we write $z = \varepsilon_0 \tilde{z}$, $g = \tilde{g}/\varepsilon_0$, and define a new variable,  $y=\varepsilon/\varepsilon_0$, with $h(y) dy =f(\varepsilon) d\varepsilon$, then (\ref{blue}) becomes
\begin{equation}
 \tilde{z}(\tilde{g}) = \frac{1}{\tilde{g}} + \int dy  h(y) \frac{y}{1 - q\tilde{g}y }, 
\end{equation}
where now $\langle y\rangle=1$.
This shows that $\varepsilon_0$ is just a scaling parameter.

For a given density $f(\varepsilon)$, the integral in  (\ref{blue}) must first be performed so as to obtain the inverse resolvent, $z(g)$, which then needs to be inverted to yield the resolvent, $z(g)$, from which one can obtain the ESD, $\rho(\lambda)$, via  (\ref{eq:rho}). For the main cases of interest here, namely when 
$f_N(\varepsilon)$  belongs to either the Wishart or the inverse Wishart classes, the integral in  (\ref{blue}) can still be carried out analytically, but its inversion  needs to be performed numerically. These two cases are discussed separately below, after we briefly consider, for completeness,  the Marchenko-Pastur law.

\subsection{Constant background: The Marchenko-Pastur distribution}
\label{sec:MP}

In the case of non-fluctuating background, we can set
 $f(\varepsilon) =\delta(\varepsilon-\varepsilon_0)$,
 so that Eq.~(\ref{blue}) becomes
\begin{equation}
\label{RMP}
    z_{\rm MP}(g) = \frac{1}{g} + \frac{\varepsilon_0 }{1 - q \varepsilon_0 g }.
\end{equation}
Solving for $g(z)$ and using (\ref{eq:rho}), one readily obtains, as expected, the Marchenko-Pastur distribution \cite{marchenko1967distribution} of eigenvalues \cite{tulino2004random, potters2020first}: 
\begin{equation}
    \rho_{\rm MP}(\lambda) = \frac{1}{2\pi q\varepsilon_0}\frac{\sqrt{(\lambda_{+}-\lambda)(\lambda-\lambda_{-})}}{\lambda},
\end{equation}
where  the ratio $q = p/T$ is within the interval $[0 ,1]$ and $\lambda_{\pm} = \varepsilon_0(1\pm\sqrt{q})^2$ are the largest and smallest eigenvalues, respectively.
Next, we analyze  the case of multiscale complex backgrounds, where the background distribution is given by either (\ref{eq:Wback}) or (\ref{eq:invWback}).

\subsection{Fluctuating hierarchical background}

We now consider the cases where the fluctuating background is described by either one of the two classes of distributions $f_N(\varepsilon)$ given in Sec.~\ref{sec:review}. Since in both cases $f_N(\varepsilon)$ is written in terms of Meijer $G$ functions, it  is  convenient first to  rewrite (\ref{blue}) as
\begin{equation} 
\label{blue2}
    z(g) = \frac{1}{g} - \frac{1}{qg} \int d\varepsilon f_N(\varepsilon) {G}_{1,1}^{1,1}\left( 
\begin{array}{l}
1 \\ 
1
\end{array}
\bigg | -  q g \varepsilon  \right),
\end{equation}
where we used that \new{\cite{mathai1991multivariate}}
\begin{equation}
\frac{x}{1-ax}=   -  \frac{1}{a}{G}_{1,1}^{1,1}\left( 
\begin{array}{c}
1\\ 
1
\end{array}
\bigg | -  ax  \right) . 
\end{equation}
The advantage of (\ref{blue2}) is that the integral can now be computed explicitly in terms of $G$ functions.
We begin with the Wishart class and then proceed to the inverse Wishart class. 

\subsubsection{Wishart Class}

For the Wishart class, $f_N(\varepsilon)$ is given by (\ref{eq:Wback}).
Inserting this expression into (\ref{blue2}) and using the convolution properties of Meijer $G$ functions \cite{de2025matrix}, we obtain
\begin{equation}
    z(g) = \frac{1}{g} - \frac{1}{qg \Gamma(\boldsymbol{\beta})}  G_{ N+1,1 } ^{ 1,N+1 }  \left( 
\begin{array}{c}
{1,{\bf 1}-\boldsymbol\beta} \\ 
{1}
\end{array}
\bigg |-\frac{g q \varepsilon_0}{\omega}  \right) ,
\end{equation}
which upon using the power absorption property of Meijer $G$ function \cite{de2025matrix} becomes
\begin{equation}
\label{wishartspectra}
        z(g) = \frac{1}{g} + \frac{\varepsilon_0}{\omega \Gamma(\boldsymbol{\beta})}  G_{ N+1,1 } ^{ 1,N+1 }  \left( 
\begin{array}{c}
{0,-\boldsymbol\beta} \\ 
{0}
\end{array}
\bigg |-\frac{g q \varepsilon_0}{\omega}  \right) 
\end{equation}
Unfortunately, Eq.~(\ref{wishartspectra}) cannot be inverted analytically for $g(z)$, but it can be solved numerically, from which $\rho(\lambda)$ can be computed from (\ref{eq:rho}); see below. However, an asymptotic calculation can be carried out (see Appendix) which allows us to determine the tail behavior of the distribution (i.e., for large $\lambda$): 
\begin{equation}
    \rho(\lambda) \propto
        \lambda^{\beta - \frac{3}{2} + \frac{1}{2N}}\exp\left[-N\beta\left(\frac{\lambda}{\varepsilon_0 q}\right)^{1/N}\right].
        \label{eq:tailrhoW}
\end{equation}
We thus see that the ESD $\rho(\lambda)$ inherits the stretched-exponential tail of the background distribution $f_N(\varepsilon)$; see (\ref{eq:tailFW}).

\subsubsection{Inverse-Wishart Class}

In this case, $f_N(\varepsilon)$ is given in (\ref{eq:invWback}), which inserted 
 into (\ref{blue2}),  and using properties of the $G$ functions, yields
\begin{equation}
\label{inversewishartspectra}
    z(g) = \frac{1}{g} + \frac{\omega \varepsilon_0}{\Gamma(\boldsymbol{\beta} + \bf{1})} G_{ 1,N+1 } ^{ N+1,1 }  \left( 
\begin{array}{c}
{0} \\ 
{0,\boldsymbol\beta}
\end{array}
\bigg |-g q \omega \varepsilon_0 \right).
\end{equation}
As before,  Eq.~(\ref{inversewishartspectra}) has to be inverted numerically to produce the resolvent $g(z)$, from which $\rho(\lambda)$ can be computed. Using the asymptotic expansion of the $G$ function \cite{sosa2018hierarchical} one can show (see Appendix) that here again $\rho(\lambda)$ has the same tail behavior of $f_N(\varepsilon)$, see (\ref{eq:tailfIW}), namely a power-law tail:
\begin{equation}
    \rho(\lambda) \propto
        \lambda^{-\beta-2}.
\label{eq:tailrhoIW}
\end{equation}

Equations (\ref{wishartspectra}) and (\ref{inversewishartspectra}) are the main theoretical results of our analysis, in that it allow us to generalize the MP distribution to include two new families of  large correlations matrices with multiple scales, namely the hierarchical Wishart and inverse Wishart classes.  We recall that setting $N=0$ in either (\ref{wishartspectra}) or (\ref{inversewishartspectra}) yields the MP distribution, as discussed in Sec.~\ref{sec:MP}. Hence these two new distributions of eigenvalues generalize the MP law to the case $N>0$ for the two possible families of background distributions discussed in Sec.~\ref{sec:review}. Unfortunately, these generalized ESD's cannot be obtained in closed form. Nevertheless, they can be computed numerically by inverting Eqs.~(\ref{wishartspectra}) and (\ref{inversewishartspectra}).  Below, we show numerical results for $\rho(\lambda)$  for both classes.
In Sec.~IV we will apply our theory  to empirical correlation matrices obtained from financial data.

\subsection{Numerical results for $\rho(\lambda)$}

Note that the hierarchical eigenvalue distributions, $\rho_N(\lambda)$, determined from (\ref{wishartspectra}) and (\ref{inversewishartspectra}) have four free parameters, namely: the shape parameter $q$ and the mean variance $\varepsilon_0$ of the random vector $\mathbf{r}_t$, which are also present in the standard Marchenko-Pastur distribution, and two new parameters represented by the number $N$ of hierarchical levels (or characteristic time scales) in the background and the parameter $\beta$ that controls the tails of the distribution $\rho_N(\lambda)$. Recall that, as discussed above, $\rho(\lambda)$ does not have an upper bound eigenvalue for $N\ge 1$, but rather displays decaying tails that extend to infinity. The two new families of distributions, namely hierarchical Wishart and inverse Wishart classes, differ with respect to the nature of the tails: The former has a stretched-exponential decay, see (\ref{eq:tailrhoW}); while the latter has a power-law tail, as shown in (\ref{eq:tailrhoIW}).  Below we illustrate the behavior of $\rho_N(\lambda)$ for both classes.

\begin{figure*}[t]
	\centering
	\subfloat[\label{fig:1a}]{\includegraphics[width=0.48\textwidth]{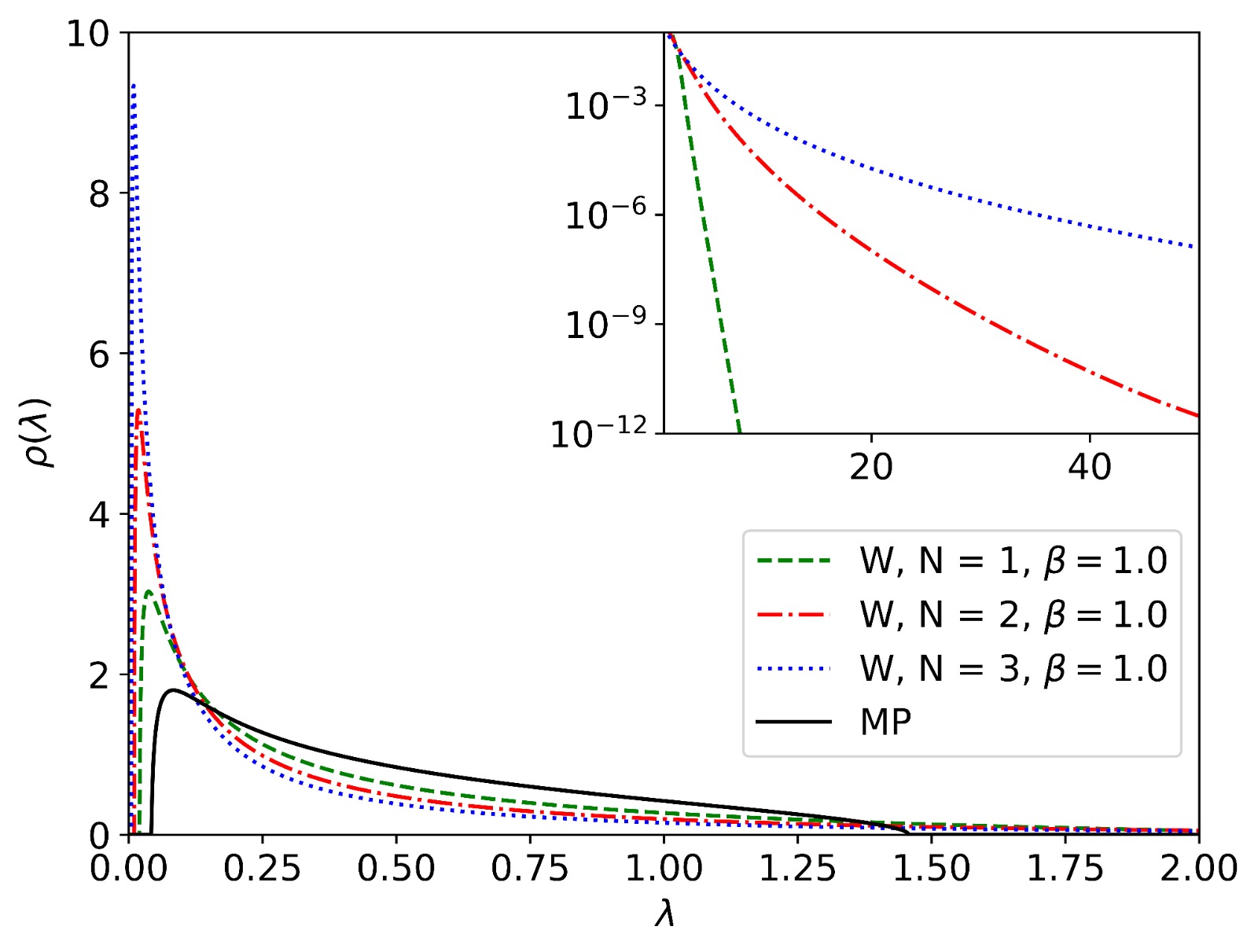}}
%	\hfill
	\subfloat[\label{fig:1b}]{\includegraphics[width=0.48\textwidth]{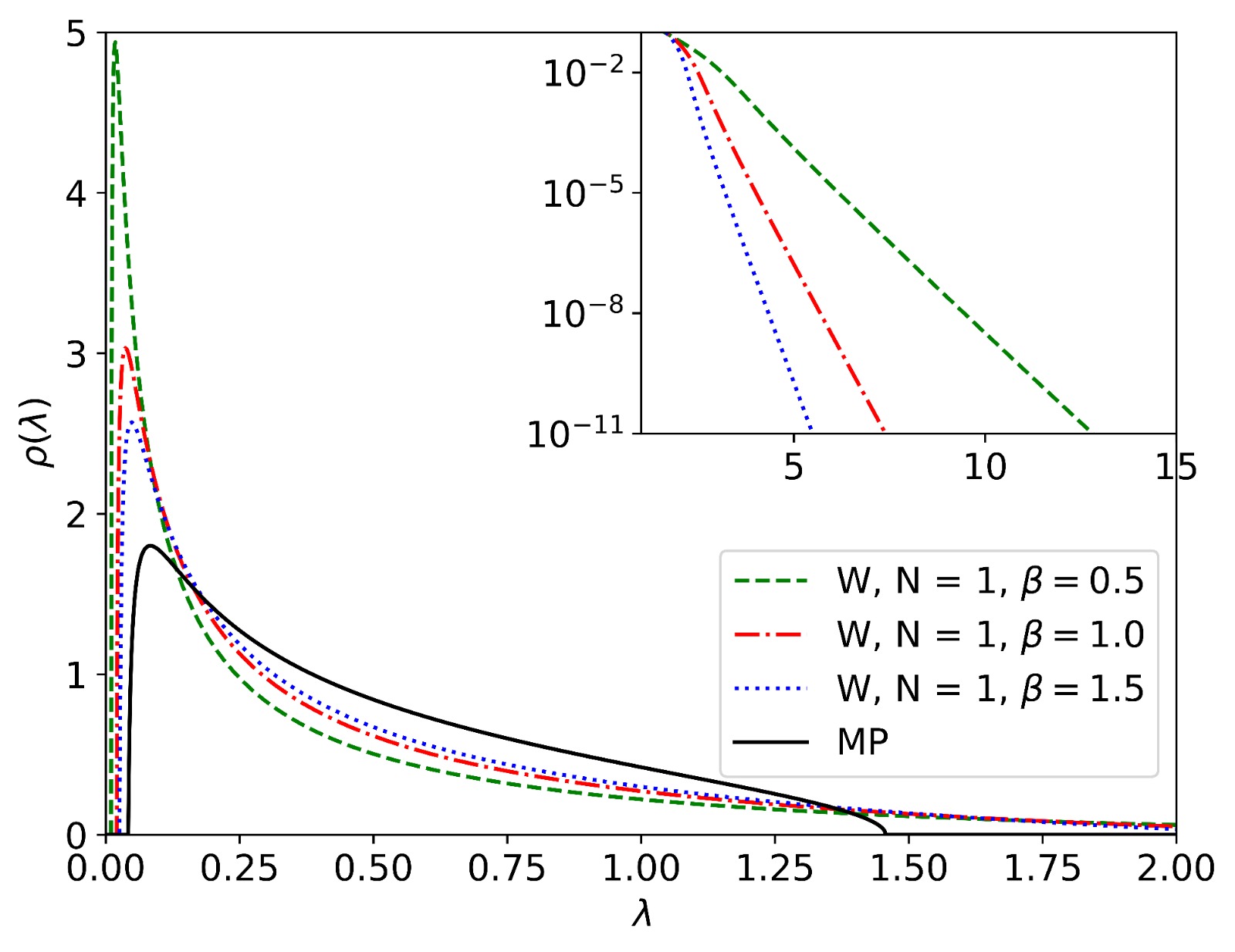}}
	\caption{Theoretical eigenvalue distributions $\rho_N(\lambda)$  for the Wishart class with fixed $q = \varepsilon_0 = 0.5$ and several values of $N$ and $\beta$. In (a) we have $\beta=  1.0 $ and $N=1, 2, 3$; whereas in (b) $N=1$ and $\beta=0.5, 1.0, 1.5$.
		The corresponding Marchenko-Pastur distribution (MP) is also shown.
		The insets show in semi-log scale the asymptotic behavior of the distributions, which is predicted by Eq.~(\ref{eq:tailrhoW}).}  
	\label{fig:W}
\end{figure*}
%
%
%\begin{figure}[t]
%    %\centering
%    \begin{subfigure}{0.48\textwidth}
%    \includegraphics[width=\textwidth]{WchangingN.png}
%    \caption{}
%    \label{fig:1a}
%  \end{subfigure}
%%  \hfill
%  \begin{subfigure}{0.48\textwidth}
%\includegraphics[width=\textwidth]{WchangingBeta.png}
%    \caption{}
%    \label{fig:1b}
%  \end{subfigure}  \caption{Theoretical eigenvalue distributions $\rho_N(\lambda)$  for the Wishart class with fixed $q = \varepsilon_0 = 0.5$ and several values of $N$ and $\beta$. In (a) we have $\beta=  1.0 $ and $N=1, 2, 3$; whereas in (b) $N=1$ and $\beta=0.5, 1.0, 1.5$.
%  The corresponding Marchenko-Pastur distribution (MP) is also shown.
%  The insets show in semi-log scale the asymptotic behavior of the distributions, which is predicted by Eq.~(\ref{eq:tailrhoW}).}  
%  \label{fig:W}
%\end{figure}

In Fig.~\ref{fig:W} we show the theoretical eigenvalue distributions, $\rho_N(\lambda)$, for the Wishart class, as obtained by inverting numerically Eq.~(\ref{wishartspectra}) and then applying (\ref{eq:rho}).  In these figures, we have kept $q$ and $\varepsilon_0$ fixed (namely, $q=\varepsilon_0=0.5)$ and have varied the new
parameters  $\beta$ and  $N$.  We also show for comparison the corresponding MP distribution (solid black line) with the same values of $q$ and $\varepsilon_0$. As one can see from the main panel of Fig.~\ref{fig:1a}, as the number, $N$, of hierarchical levels increases (for fixed $\beta$), the distribution's peak gets higher and shifted to the left, with the minimum eigenvalue becoming consequently smaller (i.e., closer to zero). A similar behavior, but in reverse order,  is observed for varying $\beta$ (with fixed $N$): increasing $\beta$ lowers the peak and shifts it to the right, as seen in Fig.~\ref{fig:1b}. 
%\new{It is thus reasonable to expect that the position of the peak should depend (at least in first approximation) mostly on the ratio  $N/\beta$.} %({\bf Será verdade isso? Eu coloquei na versão anterior mais como uma conjectura. Deixamos ou tiramos?)} 
%
In the insets of Fig.~\ref{fig:W} we display in semi-log scale the tails of the corresponding distributions shown in the main panels. As one can see, the observed tails  are in agreement with the theoretical prediction in (\ref{eq:tailrhoW}), namely: for $N=1$ the distribution has an exponential tail (hence a straight line in the semi-log scale), whereas for $N>1$ one sees a heavier tail as expected from the stretched-exponential behavior given in (\ref{eq:tailrhoW}).

\begin{figure*}[t]
	\centering
	\subfloat[\label{fig:2a}]{\includegraphics[width=0.48\textwidth]{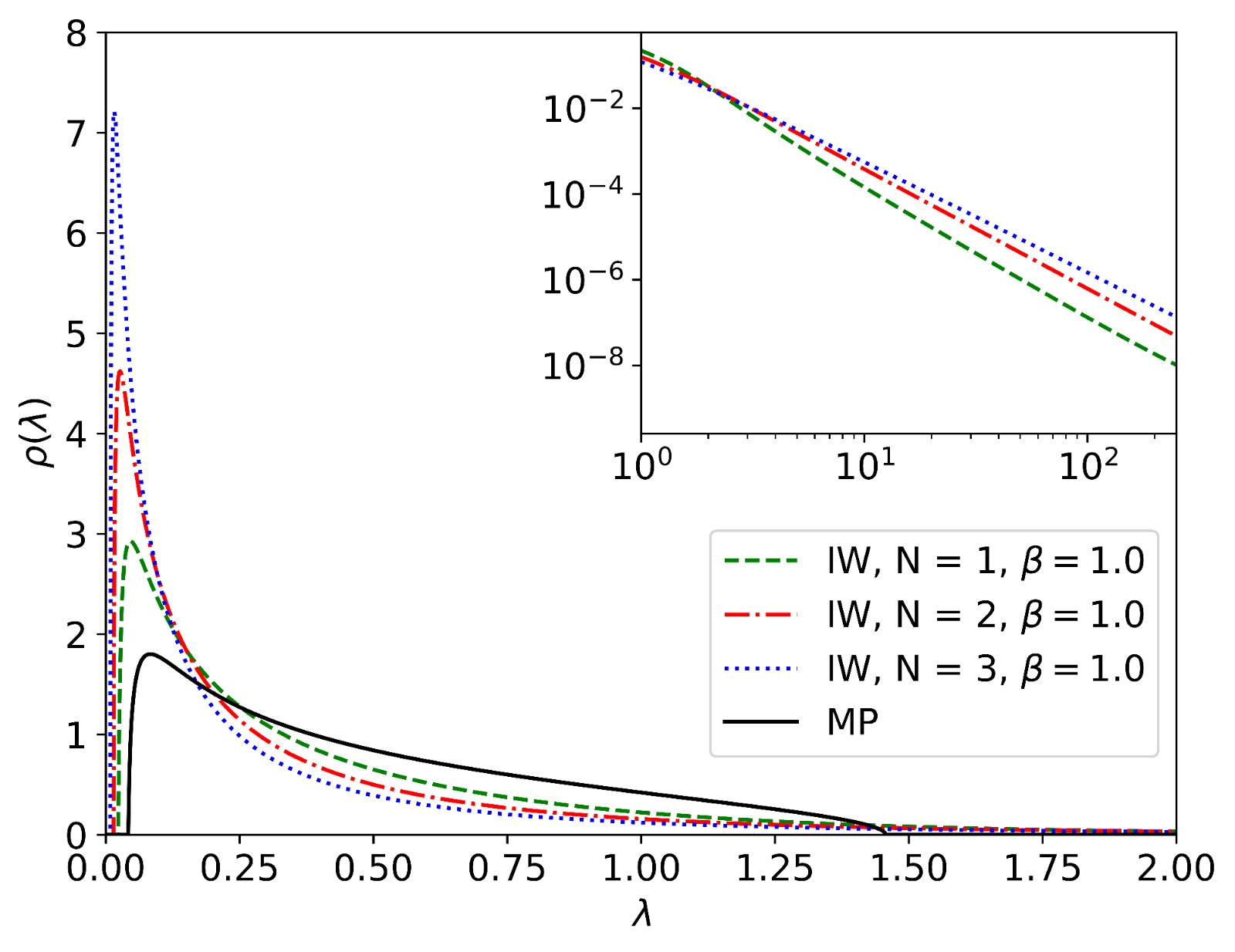}}
%	\hfill % Adiciona espaço horizontal entre as figuras
	\subfloat[\label{fig:2b}]{\includegraphics[width=0.48\textwidth]{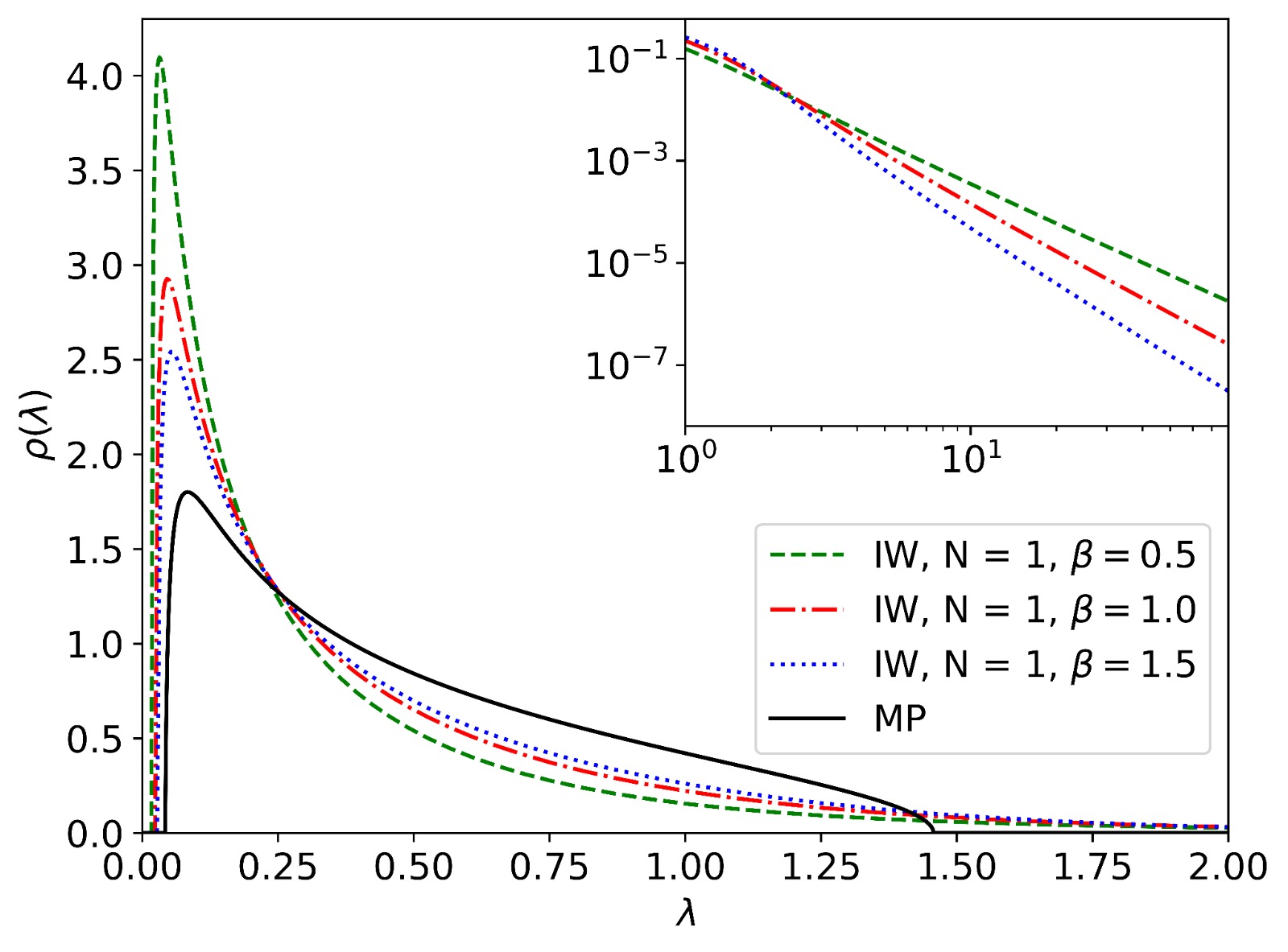}}
	\caption{Theoretical eigenvalue distribution $\rho_N(\lambda)$ for the inverse Wishart class, with fixed $q = \varepsilon_0 = 0.5$ and several values of $N$ and $\beta$. In (a) we have $\beta=  1.0 $ and $N=1, 2, 3$; whereas in (b) $N=1$ and $\beta=0.5, 1.0, 1.5$.
		The corresponding Marchenko-Pastur distribution (MP) is also shown.
		The insets show in log-log scale the power-law tails of the distributions, as predicted by Eq.~(\ref{eq:tailrhoIW}).} 
	\label{fig:IW}
\end{figure*}

%\begin{figure}[t]
%    \centering
%    \begin{subfigure}{0.49\textwidth}
%   \includegraphics[width=\textwidth]{IWchangingN.png}
%    \caption{}
%    \label{fig:2a}
%  \end{subfigure}
%  \begin{subfigure}{0.49\textwidth}
%     \includegraphics[width=\textwidth]{IWchangingBeta.png}
%    \caption{}
%    \label{fig:2b}
%  \end{subfigure}
%  \caption{Theoretical eigenvalue distribution $\rho_N(\lambda)$  for the inverse Wishart class, with fixed $q = \varepsilon_0 = 0.5$ and several values of $N$ and $\beta$. In (a) we have $\beta=  1.0 $ and $N=1, 2, 3$; whereas in (b) $N=1$ and $\beta=0.5, 1.0, 1.5$.
%  The corresponding Marchenko-Pastur distribution (MP) is also shown.
%  The insets show in log-log scale the power-law tails of the distributions, as predicted by Eq.~(\ref{eq:tailrhoIW}).} 
%  \label{fig:IW}
%\end{figure}

After having discussed the main aspects of our novel results, namely the two classes of hierarchical eigenvalue distributions,  we shall next apply our theory to model the empirical distribution of eigenvalues obtained from financial data.

\section{Application to Financial Data}
\label{applications}

According to \cite{plerou1999universal, plerou2002random, laloux1999noise} the eigenvalue spectrum of correlation matrices of financial returns in the 1990's is mostly described by the MP distribution. It suggests that most of the content of the correlation matrix is random and homoscedastic, in addition to some outliers eigenvalues \cite{plerou2002random}. This results shows that RMT-based models are suitable to describe the empirical correlation matrix. The eigenvalues that violate this description are often associated with industrial branches in the market and the market index itself. One might then argue that the eigenvalues that match the MP distribution description are related to noise -- they reveal no information about how assets correlate with each other. Here we will see, however, that this description no longer seems to be appropriate (for more recent data). More specifically,  we will show below that the empirical correlation matrix of recent financial data is not well described by a MP distribution but rather displays  a well defined tail that can be more efficiently captured by our hierarchical models described in Sec.~\ref{sec:eigen}. 

\subsection{Empirical correlation matrix}

\begin{figure}[t]
    \centering
    \includegraphics[width=1\linewidth]{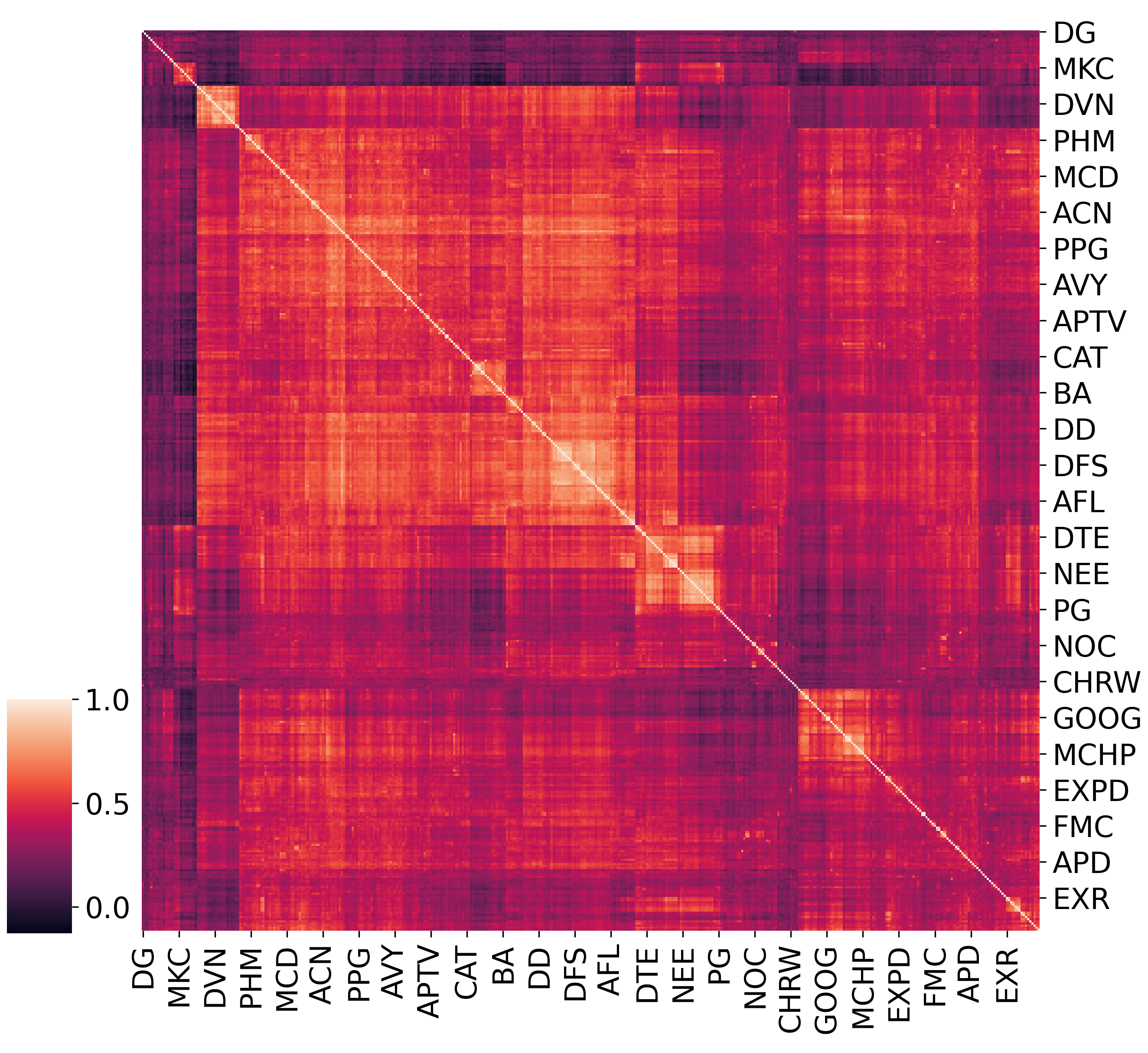}
    \caption{Empirical correlation matrix of stock returns in the S\&P 500 between 5/1/2020 to 5/1/2025. The clustering in the heatmap is obtained using the hierarchical clustering method based on the correlation distance \cite{mantegna1999introduction}. This matrix captures the observed correlations, which include both true market structure and finite-size noise.}
    \label{fig:clustermap_app}
\end{figure}

We have analyzed a dataset of  daily close prices of 424 stocks belonging to the S\&P 500 stock index, covering the period from  January 3, 2020, to January 3, 2025, totalizing 1259 trading days. Let us represent the set of the $p=424$ stock prices at given time $t$ as the vector $\mathbf{x}(t)$, where   $t=1,...,T$, with $T=1259$.   As usual, we define the price returns by $\mathbf{r}(t)$, where $ r_i(t) = \log[x_i(t+1)] - \log[x_i(t)]$. 

% \begin{figure}[t]
%     \centering
%     \includegraphics[width=1\linewidth]{MPFailure2.png}
%     \caption{Main figure: MP-Distribution with $q = 424/1259$ as a fixed parameter and $\varepsilon_0 = 0.29$, obtained using the least minimum squares. Inset: The empirical distribution of eigenvalues in large scale, showing that eigenvalues ranges from units to hundreds.}
%     \label{fig:MPFailure}
% \end{figure}
\begin{figure}[t]
    \centering
    \includegraphics[width=1\linewidth]{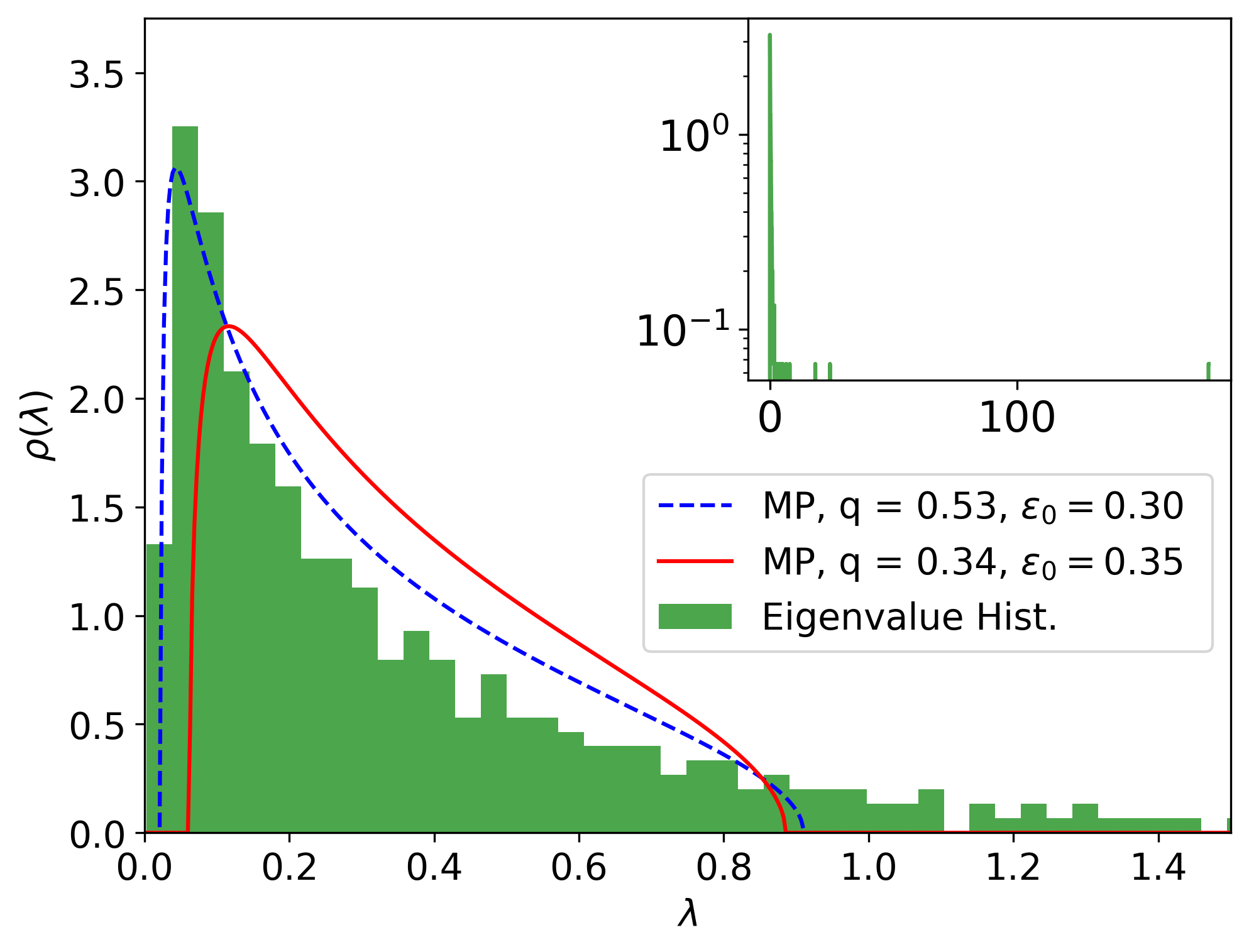}
    \caption{Main figure: MP-Distribution with $q = 424/1259$ as a fixed parameter and $\varepsilon_0 = 0.29$, obtained using the least minimum squares. Inset: The empirical distribution of eigenvalues in large scale, showing that eigenvalues ranges from units to hundreds.}
    \label{fig:MPFailure}
\end{figure}

We next calculate the eigenvalue histogram, $\rho(\lambda)$, of our correlation matrix $C$, as defined in Eq.~(\ref{eq:C_sec3}) and shown in Fig.~\ref{fig:clustermap_app}. The resulting empirical ESD (green bars) is presented  in Fig.~\ref{fig:MPFailure}. 
In this figure we also show  the best fit by the MP distribution (red solid line) with $q = p/T = 424/1259$ fixed  and obtaining  $\varepsilon_0=0.29$ as the best-fitting value. (All fitting procedures employed in this work were performed with  the least squares method.) The inset in Fig.~\ref{fig:MPFailure} shows the outliers that usually do not fit the RMT description \cite{laloux1999noise}.
We see from Fig.~\ref{fig:MPFailure} that the MP distribution provides a poor description of the empirical ESD. This is in contrast with the results of \cite{plerou1999universal, plerou2002random}, where it was shown that 
the MP distribution fits reasonably well the  eigenvalue spectra of the S\&P 500 returns during the 1990's. We note  that  if we allow $q$ to be a free parameter, the MP fit improves but it still gives a relatively poor match to the data; see blue dashed line in Fig.~\ref{fig:MPFailure}. 
The failure of the MP distribution in describing our more recent data might indicate that the empirical correlation matrix of  the S\&P 500 returns is no longer described by pure homoscedastic noise. 
Nevertheless, we expect that this failure of RMT (insofar as the MP distribution is concerned)  can be overcome by considering the emergence of additional characteristic time scales in the underlying market dynamics, which in turn can be grasped by using the hierarchical generalization of the MP distribution  predicted by matrix H theory. The second main objective of the present paper is precisely to investigate the eigenvalue spectrum of the empirical correlation matrix of the S\&P 500 index over a more recent period of time in  light of the eigenvalues distributions $\rho_N(\lambda)$ described in Sec.~\ref{sec:eigen}. We now turn to this analysis. 

\subsection{Matrix H theory analysis of return distribution}\label{HMT}

As discussed in Sec.~\ref{sec:eigen}, there are two possible families of hierarchical distributions $\rho_N(\lambda)$. Thus, for a given dataset, we first need to determine which class (Wishart or inverse Wishart) best describes the financial data.
We recall that the two hierarchical background distributions $f_N(\varepsilon)$ that define these two classes---see  Eqs.~(\ref{eq:Wback}) and (\ref{eq:invWback})---were introduced so as to yield non-Gaussian distributions of the measured signal (price returns in our case), see Eq.~(\ref{eq:univariatecompound}).   These more general distributions are required to describe the short-scale fluctuations of multiscale signals  that often display significant deviation from Gaussianity.
In order to choose the best model to describe the normalized financial returns (represented by the vector ${\bf r}_t$), we are going to rely on the procedure employed in \cite{de2025matrix}, as  described below. 

Consider the empirical correlation matrix, $C$,  given by Eq.~(\ref{eq:C_sec3}), which we assume can be diagonalized. 
Let us define the vector $\bar{\bf r}_t = U^\top {\bf r}_t$, where the matrix $U$ diagonalizes $C^{-1}$, that is,
\begin{equation}
    {\bf r}_t^\top C^{-1} {\bf r}_t = {\bf r}^\top_t U \Lambda^{-1} U^\top {\bf r}_t = \bar{\bf r}_t \Lambda^{-1} \bar{\bf r}_t
\end{equation}
where $\Lambda^{-1}$ is a diagonal matrix (namely, $C^{-1}$ in its basis of eigenvalues). After that, we define the vector $\tilde{\bf r}_t = \Lambda^{-1/2} \bar{\bf r}_t$. This procedure guarantees that all processes now are uncorrelated and normalized to unit. As discussed in detail in \cite{schmitt2013non, manolakis2023analysis, de2025matrix}, all processes, $\tilde r_i(t)$, $i=1,...,p$ and $t=1,...,T$, written in the new basis can be viewed as different realizations of the same stochastic processes described by the univariate projections described in table \ref{tab:comparison}. Since all returns in the new basis obey the same statistics and are described by a single univariate distribution, we can aggregate the multiple time series $\tilde{r}_i(t)$  into a single time series, $R(t)$, $t=1,...,p\times T$, which will be referred to as the {\it aggregated returns}. We can now analyze the aggregated returns $R(t)$ in terms of the univariate projection distributions given in (\ref{wishartaggregated}) and (\ref{inversewishartaggregated}). 

\begin{figure}
	\centering
	\includegraphics[width=1\linewidth]{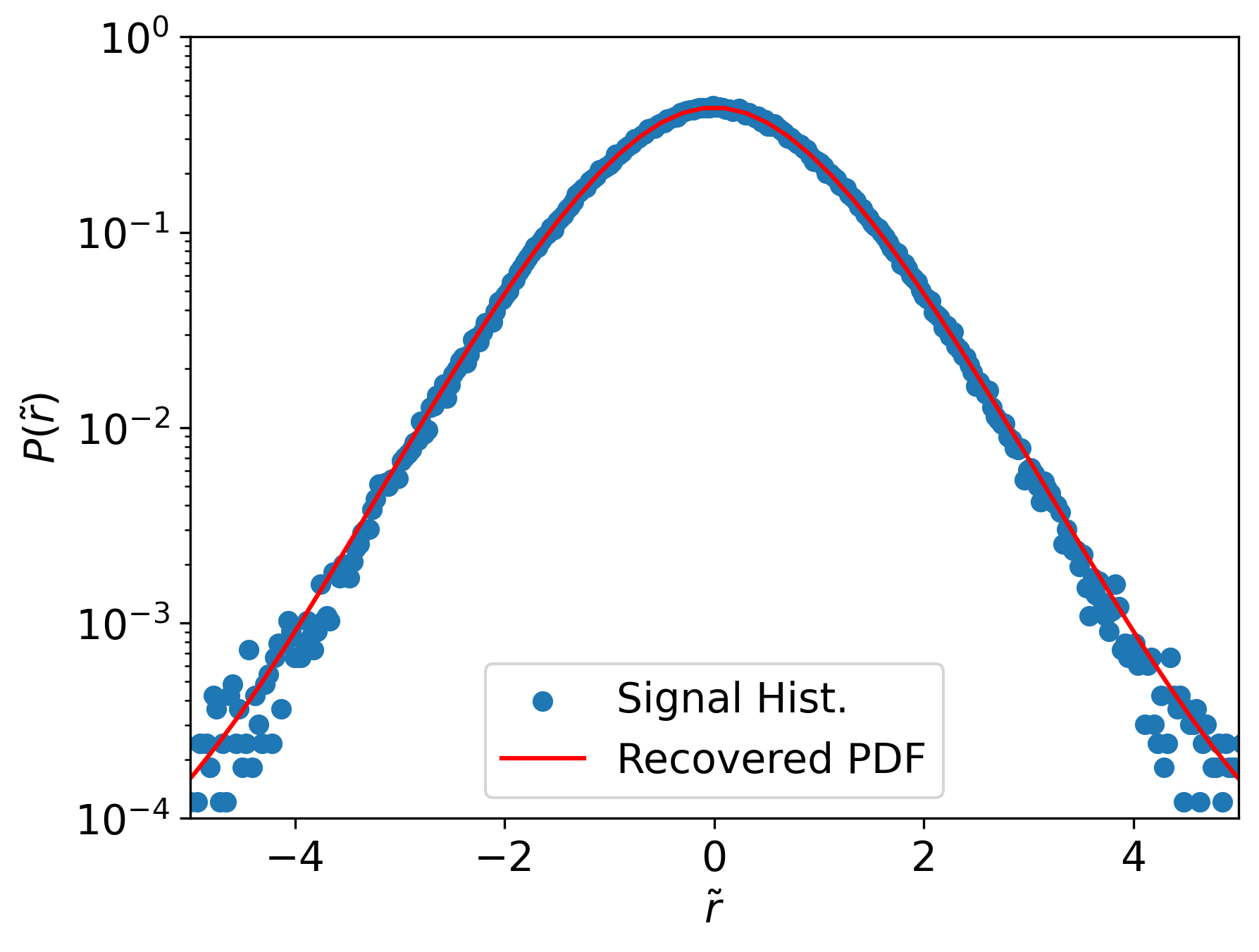}
	\caption{The aggregated signal histogram and the recovered probability density function with $L=18$ using the methods reported in Section \ref{HMT}. The recovered PDF describes nicely the aggregated histogram.}
	\label{fig:aggreturnsvsRecoveredPDF}
\end{figure}

As discussed in \cite{de2025matrix}, it is more discriminating to perform the theoretical fits at the level of the background distribution rather than try to fit directly the return distribution. To do that, we  first need to obtain the background empirical distribution from the aggregated returns  $R(t)$. For this, we use an auxiliary series of variance estimators on moving windows of size $L$: $\varepsilon_L(t) = \frac{1}{L}\sum_{j=0}^{L-1}[R(t - j\delta t) - \langle R(t)\rangle_L]^2$, where $\langle R(t)\rangle_L = \frac{1}{L}\sum_{j=0}^{L-1} R(t-j\delta t)$. In order to determine the optimal window size $L$, we  compound a normal distribution with the empirical distribution $\varepsilon_L(t)$, as described by Eq.~(\ref{eq:univariatecompound}), and compare the result with the empirical aggregated distribution of returns.
The best window size, $L^*$, is chosen as the one that minimizes the corresponding root mean square error.  For our data we found $L^* = 18$.
In Fig.~\ref{fig:aggreturnsvsRecoveredPDF} we show the empirical  distribution of aggregated returns (blue circles) on which we superimposed the compound distribution (solid red line) between the normal distribution and the empirical distribution of variances for the optimal $L^*$. As the figure shows, the compound distribution reproduces very well the empirical return distribution, thus attesting that our empirical series of variance is a reliable estimate of the fluctuating variance in our our original dataset. 

\begin{figure*}[t]
	\centering
	\subfloat[\label{Fig:WbgPlot}]{\includegraphics[width=0.48\textwidth]{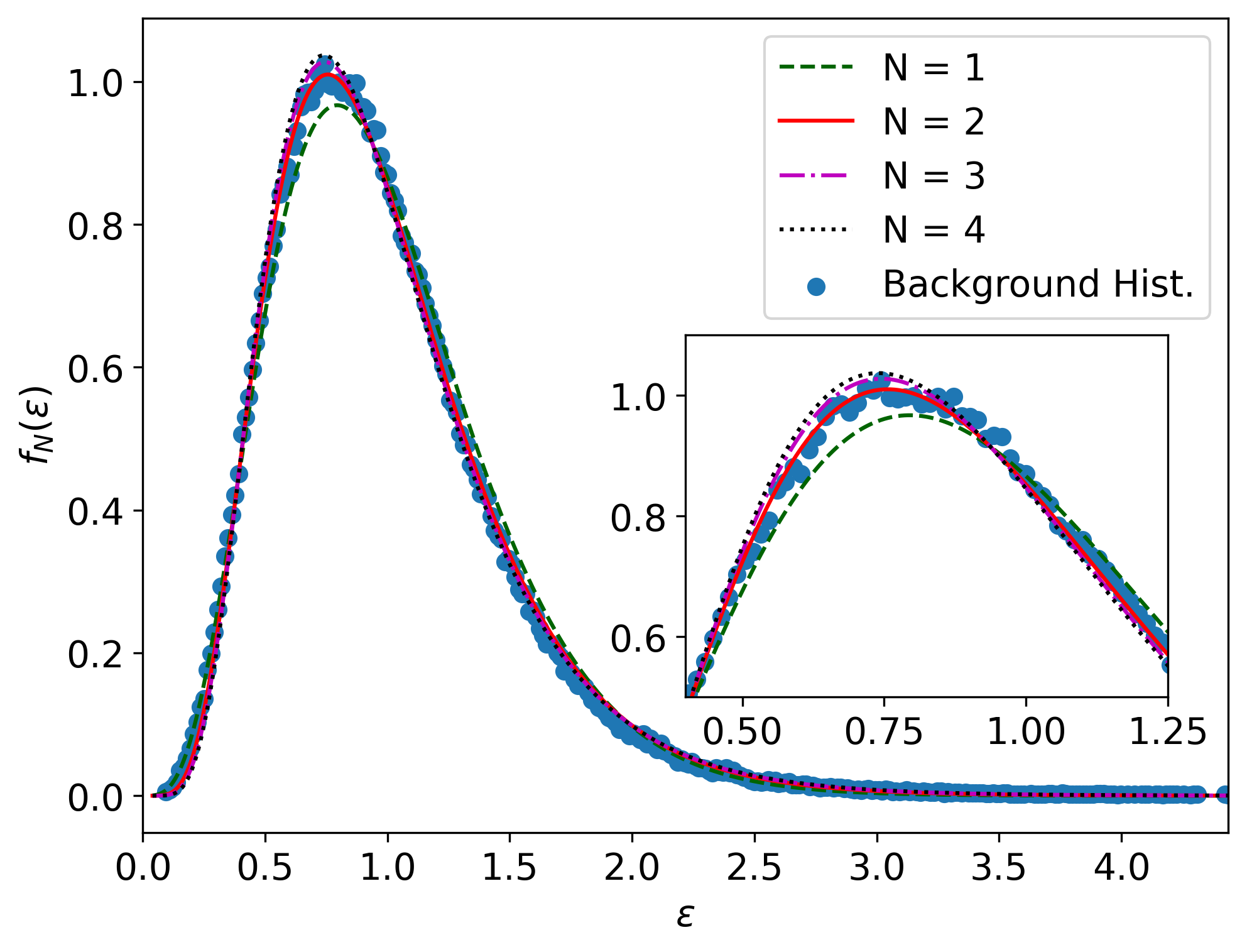}}
%	\hfill % Adiciona espaço horizontal entre as figuras
	\subfloat[\label{Fig:IWbgPlot}]{\includegraphics[width=0.48\textwidth]{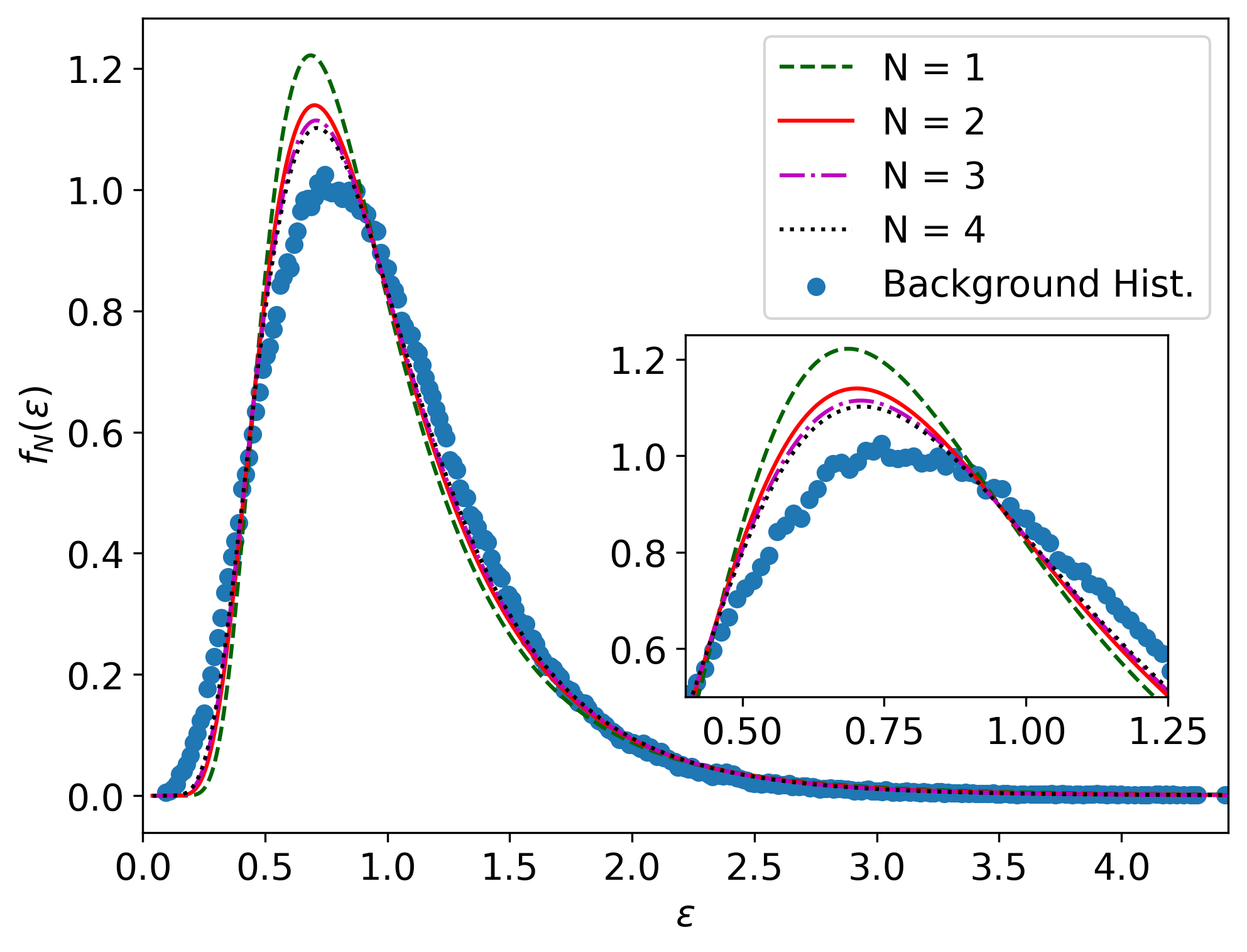}}
	\caption{Background distribution $f_N(\varepsilon)$. In (a), the Wishart class shows a good description of the background histogram. The best fitting occurs at $N=2$, as it is possible to infer from the inset, which shows a zoomed-in region around the peak, and  is confirmed by an error analysis}. In (b), the inverse Wishart class shows a much poorer fit. None of the curves seems to fit  well the histogram.
	\label{fig:back}
\end{figure*}

%\begin{figure}
%	\centering
%	\includegraphics[width=1\linewidth]{ErrorsWishartIWishart.png}
%	\caption{Values of RSS for both universality classes of H-Theory. Notice that not only Inverse-Wishart class have for all N a bigger RSS, but a clear minimum in N = 2 Wishart class. This makes possible to choose the most appropriate model to describe our data.}
%	\label{fig:RSSvsN}
%\end{figure}

With the empirical aggregated background distribution in hand, it is then possible to perform a fitting procedure of this distribution with  the two theoretical background distributions $f_N(\varepsilon_N)$, see Eqs.~(\ref{eq:Wback}) and (\ref{eq:invWback}),
in order to determine which model  best  describes the data. A similar analysis was performed in \cite{de2025matrix} for the S\&P 500 returns (albeit for a different period than the one considered here), and there it was found that the Wishart class  describes better the data than does the inverse Wishart class.  We have checked that the same applies here, as can be seen in Fig.~\ref{fig:back}, where we plot the empirical distributions of variances and the theoretical fits with various $N$ for the two classes given  in  Eqs.~(\ref{eq:Wback}) [Fig.~\ref{Fig:WbgPlot}] and  (\ref{eq:invWback}) [Fig.~\ref{Fig:IWbgPlot}]. Here, we have set  $\varepsilon_0 = 1$, since the data was normalized to unit variance, and then for each $N$ we obtained the optimal $\beta$ that minimizes the error between the theoretical curve and the empirical  background distribution. 
From Fig.~\ref{fig:back}, one already sees by visual inspection that the theoretical background distribution for the Wishart class  provides a  better description of the data than that for the Inverse-Wishart class---this is particularly noticeable in the insets of Fig.~\ref{fig:back} which show zoomed-in regions around the peaks. Indeed, an analysis of the residual errors for the fitting procedures shown in Fig.~\ref{fig:back} confirms that the minimum error occurs for the Wishart class with $N = 2$.
%\old{This is further confirmed by Fig.~\ref{fig:RSSvsN} which shows  the  fitting errors of the plots in Fig.~\ref{fig:back} as a function of $N$, where one sees that the errors for the Wishart classes are significantly lower than those for the inverse Wishart class. Furthermore, Fig.~\ref{fig:RSSvsN} reveals  an overall  minimum error for the Wishart class at $N = 2$, thus showing that this class with $N = 2$ and $\beta = 9.34$ yields the best description of the dataset considered here.}  

\begin{figure}[t]
	\centering
	\includegraphics[width=1\linewidth]{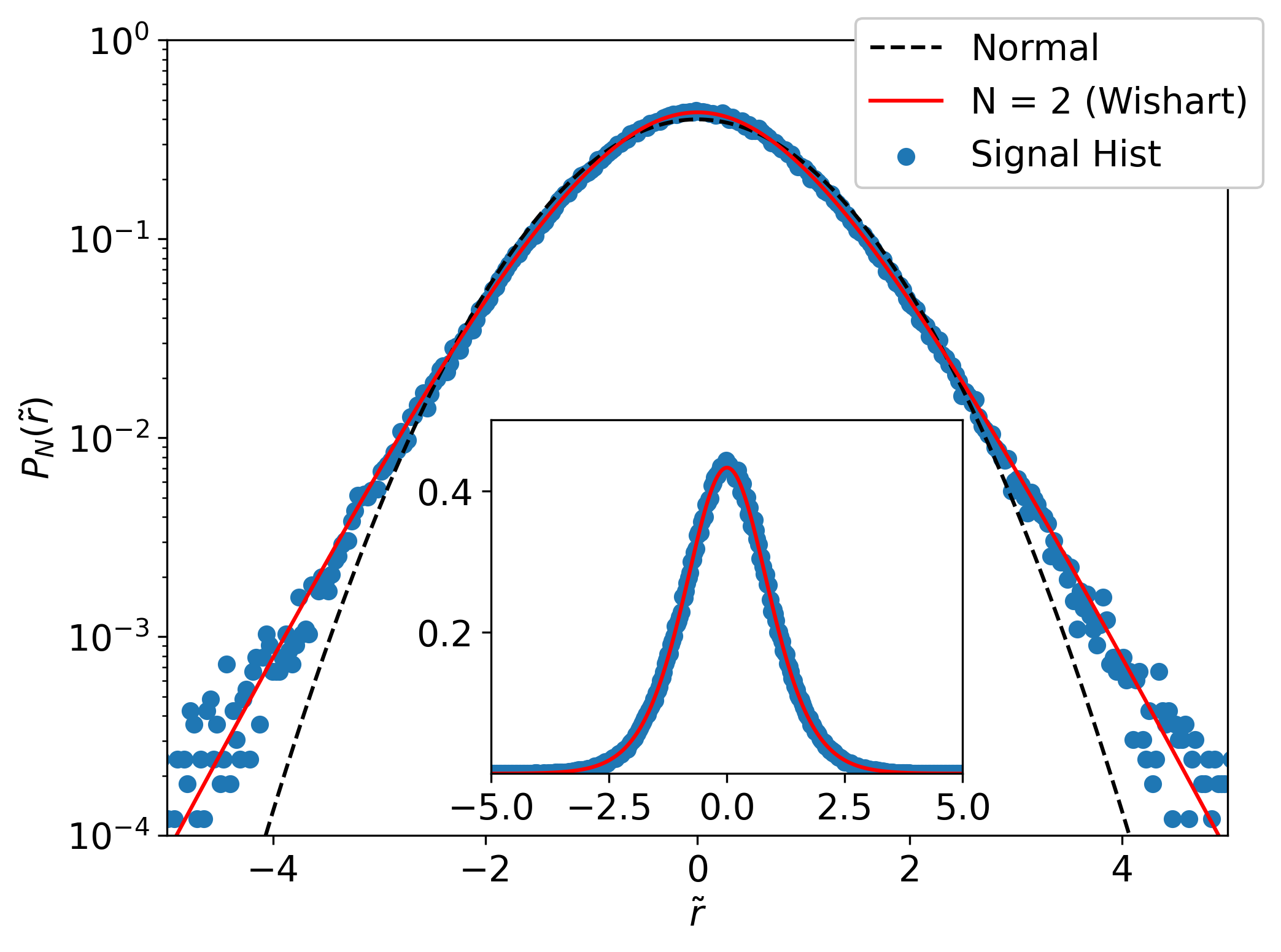}
	\caption{Aggregated signal histogram and the plot of the Eq.  (\ref{wishartaggregated}) for $N = 2$, $\beta = 9.57$ and $\varepsilon_0 = 1$. The accordance of the theoretical curve the empirical data is excellent. We also display the Gaussian distribution ($\mu = 0, \sigma 
= 1$) showing that it provides a poor fit.}
	\label{fig:signalN=2}
\end{figure}

In Fig.~\ref{fig:signalN=2} we show  the empirical distribution of returns together with the theoretical prediction given by (\ref{wishartaggregated}) for the Wishart class with the best values ($N = 2$,  $\beta = 9.34$) obtained from the fits of Fig.~\ref{Fig:WbgPlot}. We see that the theoretical curve  provides an excellent description of the aggregated financial returns. We emphasize that there is no fit in Fig.~\ref{fig:signalN=2}---rather, we simply plot the theoretical prediction, $P_N(\tilde r)$, for the return distribution with the parameters estimated from the background fit in Fig.~\ref{Fig:WbgPlot}. The excellent agreement between theory and data also at the level of the returns is further proof of the consistency of our hierarchical model.

Now that we have selected the best model for our data, namely the Wishart class with $N=2$, indicating that there are two characteristic {intermediate} time-scales in the underlying dynamics of the S\&P 500 index (for the period considered here), we shall apply the corresponding prediction for the eigenvalue distribution, $\rho_N(\lambda)$, to our empirical correlation matrix, as discussed next.

\subsection{Eigenvalue Spectrum Analysis}

\begin{figure}[t]
	\centering
	\includegraphics[width=1\linewidth]{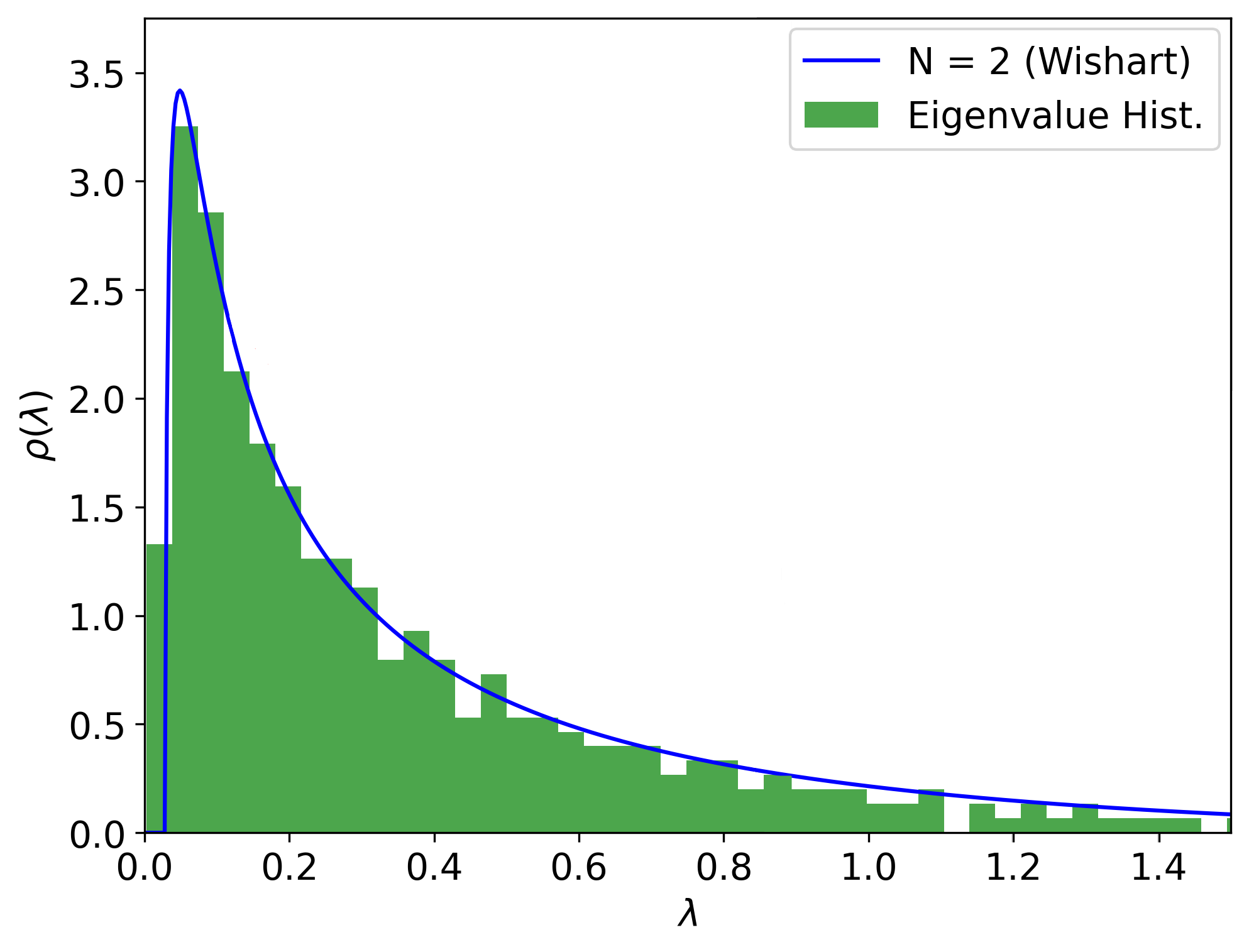}
	\caption{Empirical eigenvalue distribution and $\rho_N(\lambda)$ with $N = 2$, $q = 424/1259$, $\beta = 1.13$ and $\varepsilon_0 = 0.43$ and the MP Distribution with same parameters as reported in Fig. \ref{fig:MPFailure}. The hierarchical model provides a much better description of the eigenvalue distribution.}
	\label{fig:EigenvalueFitN=2Wishart}
\end{figure}

As described in Section \ref{sec:eigen}, in order to obtain the eigenvalue distribution $\rho_N(\lambda)$ for the Wishart class, we first need to invert Eq.~(\ref{wishartspectra}) numerically and then use (\ref{eq:rho}).   In our case, fixing $N = 2$ and $q = 424/1259$, there are two remaining  free parameters, namely $\beta$ and $\varepsilon_0$, to be inferred from the data. 
Figure \ref{fig:EigenvalueFitN=2Wishart} shows the best fit (blue line) for $\rho_N(\lambda)$ 
with $\beta = 1.13$ and $\varepsilon_0 = 0.43$,
together with the empirical ESD (green bars). It is remarkable that the hierarchical model is capable of describing very well the shape of the bulk region of the eigenvalue spectrum, including its tail, which cannot be captured by the MP distribution; compare Figs.~\ref{fig:MPFailure} and \ref{fig:EigenvalueFitN=2Wishart}. %\new{(ADICIONAR DISCUSSÂO SOBRE A POSSIBILIDADE DE MUDAR Q NA MP)}. 
This fact and the results obtained in Section \ref{HMT} provide  strong evidence that the empirical correlation matrix is dressed by a hierarchical (and possible turbulent) noise modeled by the H-theory approach, which is akin in spirit to Kolmogorov's statistical theory of turbulence \cite{vasconcelos2024turbulence}. It is also important to emphasize that  shuffling the vector returns  does not change the shape of the eigenvalue distribution. This is in agreement with our model where we assumed that the correlation matrix is build from i.i.d.~random vectors.
It is interesting to notice that while fitting the ESD, see Fig.~\ref{fig:EigenvalueFitN=2Wishart}, we obtained the optimal value $\beta = 1.13$, in Fig.~\ref{Fig:WbgPlot} we found  $\beta = 9.57$ when fitting the series of variances of the returns. This discrepancy calls for additional understanding of the role of the parameter $\beta$ in the two contexts, 
as discussed in the next section. 

\section{Discussion}\label{discussions}

\subsection{Discrepancies between $\beta$}

%As mentioned above, the values of the parameter $\beta$ are not equal in the two different analyses performed in the last section, namely return and eigenvalues distributions. 
We recall that in Sec.~\ref{applications}, by fitting the return data, we have determined that the  Wishart class with  $N = 2$ scales was the appropriate model for the dataset considered here.  
We then took this $N$ as the best value also for the eigenvalue distribution $\rho_N(\lambda)$. This is a reasonable assumption because,  within the context of the H-theory approach \cite{macedo2017universality, vasconcelos2024turbulence, de2025matrix}, $N$ indicates the number of relevant time scales in the underlying dynamics, and supposedly this dynamical structure should be preserved in both analysis (i.e., return and eigenvalue distributions). %It is  natural to ask, however, why these two procedures yield different values for the fitting parameter  $\beta$. The answer comes when we analyze both procedures in detail.

A key question arises from the different optimal values of $\beta$ obtained from the return distribution analysis ($\beta \approx 9.57$) and the eigenvalue spectrum analysis ($\beta \approx 1.13$). This discrepancy is not a contradiction but rather a reflection of the different nature of the two procedures. The return distribution analysis in Sec. \ref{HMT} involves a whitening transformation that removes all empirical correlations, projecting the entire multidimensional dataset onto a single aggregated series representative of the underlying volatility process. In contrast, the eigenvalue analysis in Sec. \ref{sec:eigen} focuses exclusively on the bulk of the eigenvalue spectrum, which represents the structure of the noise component under the assumption of a true diagonal correlation matrix $\Sigma_0$. The parameter $\beta$ controls the tail heaviness in both contexts, but since the two empirical objects being fitted—the aggregated background distribution and the eigenvalue noise bulk—are different projections of the original data, it is not expected that the effective tail parameter $\beta$ would be identical.

Furthermore,   $\varepsilon_0$ is a free parameter when fitting $\rho_N(\lambda)$, whereas it is set to unity  in the return distribution (since the series is normalized to unit variance). In the former case, $\varepsilon_0$ describes the total fraction of noise in the data set and thus can be seen as  a measure of how much noise is generated by the `turbulent' behavior of financial markets. This noise can in principle be filtered and the remaining part of the data  contains actual information about the correlations between financial returns and can be used to build efficient portfolios using Markowitz portfolio theory \cite{markowits1952portfolio}.

\subsection{Changes in market complexity over the years}

In  Ref.~\cite{laloux1999noise} Laloux {\it et al.}~conducted a research using  daily returns of the S\&P 500  during the years 1991–1996, where they verified that the MP distribution fitted very well the noisy eigenvalues, i.e., the `blob' region of the spectral density, of the  empirical correlation matrix. 
However, in more recent data this description seems no longer to apply. The emergence  after the 2000s  of financial technologies in large funds and banks and the availability of automatized orderings from retail investors have increased market complexity. Indeed, the volume of computational trading went from $15\%$ in 2003 to $85\%$ in 2012 \cite{glantz2013multi}. Moreover, many studies suggest that algotrading is responsible for changes in market volatility \cite{oyeniyi2024analyzing, Boehmer_Fong_Wu_2021} and it is often related as the main contributions of the 2010 Flash Crash \cite{flashcrash2010}. Furthermore, the effect of the informational cascade in financial markets plays a very important role in financial markets, since volatility can be viewed as a measure of information and it is realized differently for long-term and short-term traders. Long-term volatility affects short-term volatility, but not the opposite. This asymmetry causes an information flow cascade from larger to smaller scales, a hallmark of turbulence found in financial markets, as discussed in \cite{vasconcelos2024turbulence, zumbach2001heterogeneous, genccay2010asymmetry, chakrabarty2015investment, muller1997volatilities}. Therefore, this increase in complexity raises the question of whether the financial data returns correlations are still properly described by the usual noise dressing theory. 
The turbulence conjectured to exist in financial markets can be seen as a source of noise, thus changing dramatically the market dynamics.  The hierarchical methods discussed here, aimed to describe both the distributions of returns  and the ESD of its correlation matrix, thus seek to capture in an effective manner this underlying turbulent dynamics \cite{vasconcelos2024turbulence, de2025matrix}.\

\section{Conclusions}
 \label{conclusions}
 
Here we have presented a generalization of the Marchenko-Pastur distribution of eigenvalues of correlation matrices, considering a hierarchy of time-scales modeled via matrix H theory. Two new hierarchical families of eigenvalues distribution, corresponding to the two universality classes of dynamics predicted by H theory,  were derived. These  theoretical distributions  provide  new tools to study empirical correlation matrices of multivariate time series  in complex systems. As an application of our theory, we analyzed the empirical correlation matrix of the return time series of the stocks represented in the S\&P 500 index. Our analysis showed that  multiple time-scales seem to play an  important role in the price dynamics and its statistics as well as for the eigenvalue spectrum of the correlation matrix. Our findings show that the correlation matrix seems to be noise dressed as conjectured by \cite{plerou1999universal, plerou2002random, laloux1999noise} but with a more complex noise. Our study suggests that turbulence acts as a source of noise, where in our dataset, this hierarchical noise component, characterized by two distinct time scales, accounts for approximately 43\% of the total variance observed in the eigenvalue spectrum ($\varepsilon_0=0.43$). Thus, it is possible to use this information to filter the correlation matrix in order to obtain the actual correlation between assets, as a best inference of the S\&P 500 stocks correlation matrix $\Sigma_0$. 

We have also discussed the evolution of complexity in financial markets caused by the increase in algorithmic trading on market exchanges. Another question is to understand how the market evolves and how the number $N$ of relevant time-scales may change along the years. This important questions might show us that financial returns are not stationary when analyzed in very long time spans. 

Our hierarchical  model for eigenvalue spectra of correlation matrices does not limit itself to applications in finances and can be used to any complex system where one has the evolution of different correlated degrees of freedom, such as random lasers, neurons in the brain, schools of fishes, and so on. Future work will seek to apply the tools developed in the present study to other complex systems.

\appendix
\section{Derivation of Asymptotics}\label{apdx}
In order to obtain the asymptotic solution of (\ref{blue2}), consider that $g(z) \sim 1/z$ for $z \to \infty$ \cite{potters2020first}. The resolvent connects with the eigenvalues by the relation $g(\lambda - i\delta) = g_R(\lambda) + i\pi\rho(\lambda)$, where $\delta$ is a very small and positive number \cite{biroli2007student}. Thus, we conclude that $g_R(\lambda) \sim 1/\lambda$ for $\lambda \to \infty$.

Taking the imaginary part of (\ref{blue2}), we obtain
\begin{equation}
    0 = -\frac{\pi \rho}{g_R^2 +\pi^2\rho^2} + \int d\varepsilon f(\varepsilon) \Im\left( \frac{\varepsilon \varepsilon_0}{1-qg\varepsilon \varepsilon_0} \right),
\end{equation}
where we used the fact that $\delta \to 0$ on the left-hand side of the last equation. By assuming that $\rho/g_R$ goes to zero in the asymptotic limit, and using the Dirac delta representation inside the integral, we obtain
\begin{equation}
    0 = -\pi \rho + \frac{f(1/\varepsilon_0q g_R)}{q^2\varepsilon_0}
\end{equation}
which leads us to
\begin{equation}
    \rho \sim f(\lambda/\varepsilon_0q)
\end{equation}
It is easy to verify that for both universality classes the condition of $\rho/g_R\to 0$ is in fact satisfied with an appropriate restriction of the parameter $\beta$.

\nocite{*}
\bibliographystyle{apsrev4-2}

\end{document}